\documentclass[%
 reprint,
 amsmath,amssymb,
 aps,
pre
]{revtex4-2}

\usepackage{graphicx}% Include figure files
\usepackage{dcolumn}% Align table columns on decimal point
\usepackage{bm}% bold math
\usepackage{algorithmic}
\usepackage{algorithm}

\usepackage{amssymb} 

\usepackage{amsmath,lipsum}
\usepackage{bm}
\usepackage{multirow}
\usepackage{graphicx}
\usepackage{subfigure}
\usepackage{enumerate}

\newtheorem{definition}{Definition}
\newtheorem{assumption}{Assumption}
\newtheorem{theorem}{Theorem}

\newtheorem{lemma}{Lemma}

\newtheorem{remark}{\textbf{Remark}}
\newtheorem{corollary}{Corollary}
\begin{document}

\preprint{APS/123-QED}

\title{Zero-Determinant Strategy in Stochastic Stackelberg Asymmetric Security Game}% Force line breaks with \\

\author{Zhaoyang Cheng}
\thanks{chengzhaoyang@amss.ac.cn}
\affiliation{
Key Laboratory of Systems and Control, Academy of Mathematics and Systems Science, Beijing, 100190, China\\
 School of Mathematical Sciences, University of Chinese Academy of Sciences,Beijing, 100049, China}

\author{Guanpu Chen}
\thanks{chengp@amss.ac.cn}
\affiliation{
JD Explore Academy, JD.com Inc, Beijing, 100176, China}
\author{Yiguang Hong}
\thanks{yghong@iss.ac.cn}
\affiliation{
Department of Control Science and Engineering, Tongji University, Shanghai, 201804, China\\
Shanghai Research Institute for Intelligent Autonomous
Systems, Tongji University, Shanghai, 210201, China}

%\affil{School of Computer Science and Technology, Peking University, China}
%\affiliation{%
%Key Laboratory of Systems and Control, Academy of Mathematics and Systems Science, Beijing, 100190, China,\\  School of Mathematical Sciences, University of Chinese Academy of Sciences }

%\affiliation[1]{Key Laboratory of Systems and Control, Academy of Mathematics and Systems Science, Beijing, 100190, China, School of Mathematical Sciences, University of Chinese Academy of Sciences}
%\altaffiliation[Also at ]{Key Laboratory of Systems and Control, Academy of Mathematics and Systems Science, Beijing, 100190, China, School of Mathematical Sciences, University of Chinese Academy of Sciences}%Lines break automatically or can be forced with 

\begin{abstract}
In a stochastic Stackelberg asymmetric security game, the strong Stackelberg equilibrium (SSE) strategy is a popular option for the defender to get the highest utility against an attacker with the best response (BR) strategy. However, the attacker may be a boundedly rational player, who adopts a combination of the BR strategy and a fixed stubborn one. In such a condition, the SSE strategy may not maintain the defensive performance due to the stubborn element. In this paper, we focus on how the defender can adopt the unilateral-control zero-determinate (ZD) strategy to confront the boundedly rational attacker. At first, we verify the existence of ZD strategies for the defender.
We then investigate the performance of the defender's ZD strategy against a boundedly rational attacker, with a comparison of the SSE strategy. Specifically, when the attacker's strategy is close to the BR strategy, the ZD strategy admits a bounded loss for the defender compared with the SSE strategy. Conversely, when the attacker's strategy is close to the stubborn strategy, the ZD strategy can bring higher defensive performance for the defender than the SSE strategy does.

\end{abstract}

\maketitle

%\tableofcontents

\section{Introduction}

Stochastic security games attract more and more attention in many fields such
as the cyber-physical system (CPS), the unmanned aerial vehicle (UAV), and  the moving target defense (MTD) \cite{8485952,feng2017signaling,bondi2020signal,mutzari2021coalition}. The stochastic Stackelberg asymmetric security game is one of the important categories to characterize players' behaviors when the defender faces persistent threats from the attacker. As a fundamental model discussed in \cite{vorobeychik2012computing,korzhyk2011stackelberg}, the attacker tends to choose the best response (BR) strategy after observing the defender' strategy, while the defender aims to maximize its utility considering the attacker. Actually, the defender, as a leader, has an advantage in guiding the attacker’s decision, and the defender picks the optimal strategy based on predicting the attacker's BR strategy. The corresponding equilibrium is defined as the strong Stackelberg equilibrium (SSE) \cite{vorobeychik2012computing,li2019cooperation,9722864}.
%Recent works focus on the situation where the attacker chooses the best response strategies based on players' action history, where the defender has an advantage in predicting the attacker’s strategy \cite{qiu2021provably,li2019cooperation}.

However, players may not always be completely rational due to subjective or objective factors \cite{simon1990bounded,jiang2013monotonic} in practice. As a typical case, a boundedly rational attacker does not strictly adopt the BR strategy in practice. This may result from the limitation of the attacker's observation, the disturbance of the environment, or the imitative behavior of the attacker. For instance, in MTD problems, the attacker may not directly observe the certain defense strategy because of the disturbance from the administrator (defender) \cite{feng2017signaling,carvalho2014moving}. In UAV systems, the malicious UAV may not observe the location or the flight attitude of the legitimate UAV (defender) due to the obstruction in the wild \cite{bondi2020signal}. In CPS, the attacker may design a stealthy attack scheme instead of the BR strategy to avoid the fault detection of the defender \cite{8485952,chen2021distributed}.

When a boundedly rational attacker loses the ability or interest to achieve the BR strategy, it may likely turn to a fixed stubborn strategy in most cases. For example, a player prefers a stubborn strategy to avoid being induced to an unsatisfactory outcome in the CPS security \cite{la2016deceptive} and may choose a fixed credible strategy when the player cannot calculate the BR strategy timely in MTD problems \cite{nayak2016stubborn}. Against a stubborn attacker, the SSE strategy fails to be regarded as the optimal solution for the defender. 
In fact, a boundedly rational attacker may choose a mixed strategy, composed by the BR strategy and the stubborn strategy, and such boundedly rational players are common in security problems. For instance, in CPS, the attacker is hesitating between adopting a stubborn strategy or moving as a follower since it needs to consider the failure probability of its own data acquisition system to avoid potential loss \cite{sanjab2016bounded}. In UAV security problems, a UAV also faces different choices in different stages, since the UAV may lose the location of the defender when going through some complex terrains like in the forest, but fully observes the defender on plains \cite{bondi2020signal}. 
Thus, various factors, including the potential preference, inherent cognition, and available resources, make the boundedly rational attacker nonnegligible to the defender.
%Therein, where the agent selects the optimal action with a certain probability, and otherwise, selects a random or fixed action \cite{gal2016dropout}.

Clearly, due to the stubborn element within a boundedly rational attacker, the original SSE strategy may be no longer suitable for the defender.
Thus, it is important to consider other strategies to help the defender  maintain its defensive performance. Fortunately, zero-determinate (ZD) strategies provide a powerful idea to unilaterally enforce an advantageous relation between players' expected utilities, no matter what strategy the opponent selects. Proposed by \cite{press2012iterated} in iterated prisoner's dilemma (IPD), a ZD strategy means that one player can unilaterally enforce the two players' expected utilities subjected to a linear relation. Afterward, various ZD strategies have been widely studied to promote cooperation or unilaterally extortion in public goods games (PGG), human-computer interaction (HCI),  evolutionary games, etc \cite{wang2016extortion,hilbe2013evolution,govaert2020zero,hao2015extortion,engel2018single,szolnoki2014evolution,chen2022geometry}.

Besides, asymmetric matrix games are more realistic than symmetric ones, and there are some challenges to solve the asymmetric games due to the different preferences \cite{taha2020zero,mcavoy2015asymmetric}. Currently, there are not many breakthroughs by applying ZD strategies in symmetric games. For example, some works adopt the ZD strategies to persuade the service provider to cooperate in iterated data trading dilemma games \cite{sooksatra2018solving} and to deploy as special active defense strategies in IoT devices \cite{wang2019moving}. Considering the universality and importance of asymmetric games in security, it is necessary to explore the performance of ZD strategies in asymmetric games under security scenarios, since the original analysis of ZD strategies in IPD cannot be directly applied to the asymmetric security game.
%the application of ZD strategies in asymmetric stochastic security games still has some challanges to overcome such as the exsitence of ZD strategies and , there will be many bottlenecks to overcome in directly applying the ZD strategy method against boundedly rational attackers. Existing ZDS do not have many asymmetric results. For example,  in iterated data trading dilemma game, the user can adopt ZD strategies to unilaterally persuade the service provider to cooperate \cite{sooksatra2018solving}. In CPS, a special ZD strategy can be deployed as an active defense strategy in IoT devices \cite{wang2019moving}. However, most existent works of ZD strategies in asymmetric security games have paid less attention to the attacker's preference, especially the attacker's bounded rationality. 

%Also, the most exiting works mainly focus on promoting cooperation and pay less attention to imporve player's own reward  by utilizing asymmetric preferences.

% , the defender should also consider the rational attacker when implementing strategies, while the characteristics of the opponent are often not considered in the current ZD strategy analysis. Thus, in the asymmetric security game, the benefits of ZD strategy to the defender are still worth exploring.

In this paper, we are inspired to reveal whether the defender can adopt the ZD strategy against a boundedly rational attacker in stochastic Stackelberg asymmetric security games, in order to make up for SSE strategies' deficiencies. To this end, we show that the ZD strategy gives a better performance than the SSE strategy does. The main contribution of this work is summarized as follows.

\begin{itemize}

\item We apply ZD strategies in asymmetric security games.
We verify the existence of ZD strategies, in order to ensure the availability of the defender to adopt a ZD strategy. Besides, against the two special attackers, we investigate the defensive performance of ZD strategies compared with SSE strategies. 
Specifically, against an attacker with the BR strategy, the ZD strategy
admits a bounded loss in the utility compared with the  SSE strategy, while against a stubborn attacker, the ZD strategy performs well and brings the defender a higher utility than the SSE strategy does. 

\item We further analyze a general case where the boundedly rational attacker adopts mixed strategies. The extension takes
on analogous  tolerable results. When the attacker's strategy is close to the BR strategy, we provide the defender with appropriate ZD strategies to maintain a bounded loss in  defensive performance  compared with the SSE strategy, and save the computing resources. Also, when the attacker is close to a stubborn attacker, we show suitable ZD strategies for the defender to get higher defensive performance than SSE strategies.

\item We verify our results in two experiments by providing the defender with proper ZD strategies to compare with an SSE strategy \cite{vorobeychik2012computing}. First, we show its performance in MTD problems, where the boundedly rational attacker can directly observe the defender's strategy and derive its explicit BR strategy \cite{feng2017signaling,carvalho2014moving}. Then we show its performance in CPS problems. The setting is more complicated but practical than the considered MTD problems, where the attacker can only observe players' action history and calculate the BR strategy based on certain mechanisms, like the fictitious play and the Q-learning \cite{qiu2021provably,li2019cooperation}.

%we also show the performance of ZD strategies in the CPS compared with SSE strategies. The considered setting is complicated but realistic than the previous one in MTD, briefly speaking, where the attacker can only obser XX obtains the BR strategy based on 

\end{itemize}

\section{Stochastic Stackelberg Asymmetric Security Game}
It is known that, in a stochastic asymmetric security game with the memory of the last stage, an attacker aims to invade two targets and a defender prevents the attack in each stage \cite{guo2019inducibility,feng2017signaling}.  %The following assumption was broadly used in the literature \cite{korzhyk2011stackelberg,ijcai2019-75,ijcai2017-516}, because the defender usually tends to resist attacks and the attacker tends to implement invasions on the unprotected target. 
Consider the stochastic Stackelberg asymmetric security game $\mathcal{G}=\{\mathcal{S},N,\mathcal{D},\mathcal{A},r,P\}$.  $\mathcal{S}=\{11,12,21,22\}$ is the set of states, which is composed by the previous attack and defense targets. $N=\{d,a\}$ is the set of players. $\mathcal{D}=\{1,2\}$ and $\mathcal{A}=\{1,2\}$ are the defender's action set and the attacker's action set, respectively. $\mathbf{r}=\{r_d,r_a\}$ is the reward set of players, where $r_i:\mathcal{D}\times\mathcal{A}\to \mathbb{R}$, $i\in N$. Besides, $P:\mathcal{S}\times \mathcal{S\times\mathcal{D}\times\mathcal{A}}\to[0,1]$ is the transition function, where $P(s'|s,d,a)$ shows the probability  to the next state $s'\in\mathcal{S}$ from the current state $s$ when players take $d,a$, and $\sum\limits_{s'\in\mathcal{S}}P(s'|s,d,a)=1$ for $s\in\mathcal{S},d\in\mathcal{D}$, and $a\in \mathcal{A}$. 

In this security game, since the state presents for the previous players' actions, $P(s'|s,d,a)=1$ if and only if $s'=(da)$ for any $s\in\mathcal{S}$. Thus, the next state depends on players' strategies and the current state.  For convenience, denote $P(s'|s)$ as the state transition probability to state $s'$ from state $s$, where $s',s\in\mathcal{S}$. Furthermore, in the game $\mathcal{G}$,  each player's strategy depends on the current state, which is also a memory-one strategy. The strategy of the defender is a probability distribution $\pi_d$, where $\pi_d(d|s)\in\Delta\mathcal{D}$ with $\Delta\mathcal{D}$ denoting a probability simplex defined on the space $\mathcal{D}$. Similarly, the strategy of the attacker is $\pi_a$ with $\pi_a(a|s)\in\Delta\mathcal{A}$. Thus, $P(s,s')=\pi_d(d|s)\pi_a(a|s)$, where $s'=(da)$. Set $M=\{P(s|s')\}_{s,s'\in\mathcal{S}}$  as the state transition matrix of this security game.  As discussed in \cite{akin2016iterated,press2012iterated}, we carry forward the investigation with a regular matrix $M$.

\begin{table}[h]
\renewcommand\arraystretch{1.3}
\tabcolsep=0.2cm
\caption{Utility matrix}  
\centering  

       \begin{tabular}{cccc}
\hline
& &\multicolumn{2}{c}{Attacker} \\
& &1&2 \\
\hline 
\multirow{2}*{Defender}& 1 & $(U^d_{11},U^a_{11})$ & $(U^d_{12},U^a_{12})$ \\
~&2 & $(U^d_{21},U^a_{21})$ & $(U^d_{22},U^a_{22})$ \\
\hline
\end{tabular} 
       \label{tab::PD}  
\end{table}

At stage $t$ in $\mathcal{G}$, each player observes the current state $s_{t}$, and adopts an action according to its strategy. The defender chooses an action $d_t\in \mathcal{D}$, while the attacker chooses an action  $a_t\in \mathcal{A}$.  The reward of the the defender in stage $t$ is denoted by 
% The defender chooses an action $d_t\in \mathcal{D}$ with the probability  $\pi_d(d_t|s_{t})$, and the attacker chooses an action  $a_t\in \mathcal{A}$ with the probability  $\pi_a(a_t|s_{t})$.  The rewards of the defender and the attacker in stage $t$ are denoted by 
$
r_d(d_t,a_t)=U^d_{d_ta_t}$, where $U^d_{d_ta_t}$ is the defender's utility when the  defender protects target $d_t$ and the attacker invades target $a_t$. Similarly, the reward of the attacker in stage $t$ is denoted by $ r_a(d_t,a_t)=U^a_{d_ta_t}
$. The utility martix in each stage is shown Table \ref{tab::PD}.

\iffalse
As shown in Table \ref{tab::PD},  
$U^d_{ij}$ and $U^a_{ij}$ denote the defender's utility and the attacker's utility in each stage, 
 when the  defender protects target $i$ and the attacker invades target $j$, respectively. 
Denote $U^d=\left[\begin{array}{ll} 
U^d_{11} & U^d_{12}\\
U^d_{21} & U^d_{22}\end{array}\right]$, $U^a=\left[\begin{array}{ll}
U^a_{11} & U^a_{12}\\
U^a_{21} & U^a_{22}\end{array}\right]$, where the payoff matrix of the security game is not symmetric, i.e., $U^d\neq (U^a)^T$.

\fi

\iffalse

 The rewards of the the defender in stage $t$ are denoted by 
% The defender chooses an action $d_t\in \mathcal{D}$ with the probability  $\pi_d(d_t|s_{t})$, and the attacker chooses an action  $a_t\in \mathcal{A}$ with the probability  $\pi_a(a_t|s_{t})$.  The rewards of the defender and the attacker in stage $t$ are denoted by 
$
r_d(d_t,a_t)=U^d_{d_ta_t}$ and $ r_a(d_t,a_t)=U^a_{d_ta_t},
$ respectively, where $U^d_{d_ta_t}$ and $U^a_{d_ta_t}$ denote the defender's utility and the attacker's utility in each stage, 
 when the  defender protects target $d_t$ and the attacker invades target $a_t$, respectively. 

\fi

 % The next state $s_{t+1}=(a_tb_t)$. 
The expected long-term utilities in the repeated security game are denoted by  
$$
\begin{aligned}
U_d(\pi_d,\pi_a)&=\mathbb{E}\left(\lim_{T\to\infty}\sum\limits_{t=0}^T\frac{r_d(d_t,a_t)}{T}\right),\\
U_a(\pi_d,\pi_a)&=\mathbb{E}\left(\lim_{T\to\infty}\sum\limits_{t=0}^T\frac{r_a(d_t,a_t)}{T}\right),
\end{aligned}
$$
where  $\left\{d_{t}\sim \pi_d(\cdot|s_{t})\right\}_{t\geqslant 0}$, $\left\{a_{t}\sim \pi_a(\cdot|s_{t})\right\}_{t\geqslant 0}$, and $\left\{s_{t}\sim P(\cdot|s_{t-1},d_{t-1},a_{t-1})\right\}_{t>0}$ describe the evolution of states and actions over stage. Additionally, $s_0$ is the initial state ramdomly samplied from $\mathcal{S}$, and the expected utility of each player is the same for any $s_0$ since $M$ is convergent. %Each player aims to maximize its own expected utility. 

\begin{assumption}\label{as::1}
The utilities satisfy 
$\min\{U_{11}^d,U_{22}^d\}>\max\{U_{12}^d,U_{21}^d\}$. Moreover, $U_{11}^a<U_{12}^a$, and $U_{22}^a<U_{21}^a$.
\end{assumption}

Different from IPD \cite{press2012iterated,mamiya2020zero}, the asymmetry in this security game comes from the actual security mechanism. Specifically, the defender tends to resist attacks, that is, to protect the vulnerable target, and the attacker tends to implement invasions on the unprotected target. The above represents a wide class of asymmetric game in security scenarios, which is summarized as Assumption \ref{as::1}. Similar investigations have been broadly discussed  in the literature of various security games  \cite{korzhyk2011stackelberg,ijcai2019-75,ijcai2017-516}.

\section{Boundedly Rational Attacker}

In the stochastic Stackelberg asymmetric security game, the defender is a leader and declares a strategy in advance, while the attacker is a follower and chooses its strategy after observing the defender's strategy. In most cases, the attacker may choose the BR strategy when it obtains the defender's strategy.

%the attacker is usually a bounded rational player since it may not always strictly maximize its utility or has limited observability \cite{pita2010robust}. We consider a kind of boundedly rational attackers which chooses a mixed strategy composed by the best response (BR) strategy and  the stubborn strategy. 

%\subsection{Best Response Strategy}

After observing the defender's strategy $\pi_d$, the attacker may choose the BR strategy  \cite{li2019cooperation}  as follows:
 $$\pi_a^{BR}(\pi_d)\in \textbf{{BR}}(\pi_d)=\mathop{\text{argmax}}\limits_{\pi_a\in\Delta \mathcal{B}} U_a(\pi_d,\pi_a).$$ Without loss of
generality, the follower can break ties optimally for the leader
if there are multiple options. In this case, the defender aims to maximize
its utility considering the  attacker, and the equilibrium is defined as 
 the strong Stackelberg equilibrium (SSE) \cite{vorobeychik2012computing,li2019cooperation,9722864}.
\begin{definition}
A strategy profile $(\pi_d^{SSE},\pi_a^{SSE})$ is said to be a SSE of $\mathcal{G}$ if
$$
\begin{aligned}
&(\pi_d^{SSE}, \pi_a^{SSE})\in\mathop{\emph{argmax}}\limits_{\pi_d, \pi_a\in {\textbf{\emph{BR}}}({\pi_d})}U_d(\pi_d,\pi_a).
\end{aligned}
$$
\end{definition}

When  the attacker chooses the BR strategy after observing the defender's strategy, the SSE strategy $\pi_d^{SSE}$ is optimal for the defender, and the defender has an advantage in guiding the attacker's strategy decision. Besides, if the attacker only observes players' action history instead of the defender's strategy directly, the attacker can also choose the BR strategy by some methods such as the fictitious play and the Q-learning method \cite{qiu2021provably,li2019cooperation}.

However, the attacker may not always choose the BR strategy in security problems, due to subjective or objective factors such as the limitation of the attacker’s observation, the disturbance of the environment, and the imitative behavior of the attacker \cite{8485952,feng2017signaling,carvalho2014moving}.
In practice, the attacker may turn to other strategies. A fixed stubborn strategy, which is not influenced by the defender, is one of the most likely options for the attacker due to its potential preference, inherent cognition, and available resources \cite{la2016deceptive,nayak2016stubborn}.% Next, we import the stubborn strategy as follows.

Denote the stubborn strategy in this security game by $\pi_a^*$, while the corresponding attacker is actually a stubborn player.
In fact, the attacker intends to keep its action once it finds the most attractive target. For instance, in MTD, there always exists the most vulnerable target for the hacker, and the hacker has no intention to change its attack target once it finds the target \cite{zhuang2014towards}. Besides, a UAV tends to keep attacking the current optimum target when it has a limited vision and lacks resources to detect others \cite{shan2018modeling}. Without loss of generality, we consider that there exists a target which is more attractive than  the other for the attacker, and summarize the above in the following assumption, which was also broadly considered in \cite{la2016deceptive,zhuang2014towards,shan2018modeling}.

\iffalse
Here, we denote the stubbron strategy in this game as $\pi_a^*$.
In fact, it is not a bad choice to keep its attack target for the attacker when it does not follow the defender, that is, XXX. For instance, in MTD problems, there always exists the most vulnerable target for the hacker, and the hacker has no intention to change its attack target once it finds the target \cite{zhuang2014towards}. Besides, in UAV security problems, a UAV tends to keep attacking the current optimum target when it has a limited vision and lacks resources to detect others \cite{shan2018modeling}. Without loss of generality, we consider that there exists a target which is more attractive than  the other for the attacker, and the following assumption was widely considered in \cite{la2016deceptive,zhuang2014towards,shan2018modeling}, where the attacker intends to keep its action once it finds the most attractive target.
\fi

\begin{assumption}\label{as::pi}
Target $1$ is more attractive than  target $2$ for the attacker, i.e., $\pi_a^*(1|11)=\pi_a^*(1|21)=1$.
\end{assumption} 

%Assumption \ref{as::pi} was widely considered in \cite{la2016deceptive,zhuang2014towards,shan2018modeling}, where the attacker intends to keep its action once it finds the most attractive target.

In fact, either the BR strategy or the stubborn strategy may not be the single optimal option for the attacker. The attacker may adopt a mixed strategy composed by  both strategies. The attacker, in this case, is actually called a boundedly rational player, and is not unusual in reality. For instance, the attacker may be hesitating between BR strategies and stubborn strategies due to the errors of the data acquisition system in MTD, and data missing in UAV \cite{sanjab2016bounded,bondi2020signal,gal2016dropout}. Thus, we formulate the mixed strategy as follows.

\begin{figure*}[htp]
\begin{equation}
\begin{split}\label{eq::Gamma1}
&\Gamma_1=\{\lambda\in[0,1]|(U_{d}^{SSE}-U_{12}^d)D(\textbf{1})\lambda^4- A(1-\lambda)^4-B\lambda(1-\lambda)\geqslant 0\},\\ 
&\Gamma_2=\{\lambda\in[0,1]|(U_{12}^d-U_d^{SSE})D(\textbf{1})\lambda^4+ A(1-\lambda)^4-B\lambda(1-\lambda)\geqslant 0\}, 
\end{split}
\end{equation}
\begin{align}\label{eq::H}
&H(\pi_d^{ZD},\pi_d^{SSE},\pi_a^*,\lambda)=\frac{\left(U_d^{SSE}-U_{12}^d\right)C(\pi_d^{ZD},\pi_d^{SSE},\pi_a^*,1)\lambda^4 - A(1-\lambda)^4+B\lambda(1-\lambda)}{C(\pi_d^{ZD},\pi_d^{SSE},\pi_a^*,\lambda)}.
\end{align}
\end{figure*}

%\subsection{Mixed Strategy}

\begin{definition}\label{def::2}
The attacker's strategy  $\pi_a^{\lambda}(\pi_d,\pi_a^*)$ is called a boundedly rational strategy if
$$
\pi_a^{\lambda}(\pi_d,\pi_a^*)=\lambda \pi_a^{BR}(\pi_d) + (1-\lambda) \pi_a^*, \lambda\in[0,1].
$$
\end{definition}

For the defender's strategy $\pi_d$, we consider that the boundedly rational attacker adopts the BR strategy  $\pi_a^{BR}(\pi_d)\in \mathbf{BR}(\pi_d)$ with probability $\lambda$ and the stubborn strategy $\pi_a^*$ with probability $1-\lambda$ \cite{zychowski2021learning,jiang2013monotonic}.
\iffalse Specifically,
\begin{equation}\label{eq::mixedstrategy}
\pi_a^{\lambda}(\pi_d,\pi_a^*)=\lambda \pi_a^{BR}(\pi_d) + (1-\lambda) \pi_a^*, \lambda\in[0,1].
\end{equation} \fi
Therefore, when  the attacker selects the stubborn strategy, the defender loses the advantage in guiding the attacker's strategy decision, and the SSE strategy may not maintain the defensive performance due to the stubborn elements therein.  Thus, the SSE strategy is no longer suitable for the defender against a boundedly rational attacker.
%. If the defender still adopts an SSE strategy, the defender may obtain a lower defensive performance than its expectation, since the attacker may not choose the best response strategy guided by the defender.
It is important to study other strategies to help the defender maintain its defensive performance.

\section{Performance of ZD Strategy}
In this section, we introduce the ZD strategy for the defender in the stochastic Stackelberg asymmetric security game. At first, we show the definition of the ZD strategy for the defender and analyze the existence of the ZD strategy. Besides, we explore the performance of the ZD strategy compared with the SSE strategy.

\subsection{ZD Strategy for the Defender}

Proposed by \cite{press2012iterated}, ZD strategies mean that one player can unilaterally enforce the two players' expected utilities subjected to a linear relation, which have been widely studied to promote cooperation or unilaterally extortion in public goods game (PGG), human-computer interaction (HCI), and evolutionary games \cite{wang2016extortion,hilbe2013evolution,govaert2020zero,hao2015extortion}.
%Press and Dyson \cite{press2012iterated} proposed zero-determinant (ZD) strategies, where the player equipped with ZD strategies can unilaterally enforce the two players' expected utilities subjected to a linear relation. Afterward, various ZD strategies were widely studied to promote cooperation or unilaterally extortion in public goods game (PGG), human-computer interaction (HCI) and evolutionary games \cite{wang2016extortion,hilbe2013evolution,govaert2020zero}.
For this stochastic Stackelberg asymmetric security game $G$, the defender's ZD strategy  \cite{feng2017signaling,press2012iterated,mamiya2020zero} is defined as follows:
\begin{definition}
The defender's strategy $\pi_d^{ZD}$ is called a ZD strategy if 
\begin{equation}\label{eq::ZD-definitoin}
\begin{aligned}
\pi_d^{ZD}(1)&=\eta \mathbf{S}^{d}  +\beta \mathbf{S}^a +\gamma \mathbf{1}_4+\hat{\pi},\\
\pi_d^{ZD}(2)&=1-\pi^{ZD}_d(1),
\end{aligned}
\end{equation}
where $\eta,\beta,\gamma\in\mathbb{R}$ and $\hat{\pi}=[1,1,0,0]^T$.
\end{definition}

Let $\pi_d^{ZD}\!(k)\!\!\!=\!\!\![\pi_d(\!k|11),\!\pi_d(\!k|12),\!\pi_d(\!k|21),\!\pi_d(\!k|22)]^T\!,$ 
$k\in1,2$ and $\mathbf{S}^l=[U^l_{11},U^l_{12},U^l_{21},U^l_{22}]^T$, $l\in\{d,a\}$. The defender's all feasible ZD strategies are denoted as the
following set
$$\begin{aligned}\Xi=\{\pi^{ZD}_d\in \Delta \mathcal{A}|&\pi^{ZD}_d(1)=\eta \mathbf{S}^{d}  +\beta \mathbf{S}^a +\gamma \mathbf{1}_4+\hat{\pi},\\
& \pi_d^{ZD}(2)=1-\pi_d^{ZD}(1),\eta,\beta,\gamma\in\mathbb{R} \}.
\end{aligned}$$
It is called zero-determinant (ZD) that, if the defender adopts the ZD strategy with (\ref{eq::ZD-definitoin}), then players' expected utilities  are subjected to a linear relation:  $$\eta U_d(\pi_d^{ZD},\pi_a)+\beta U_a(\pi_d^{ZD},\pi_a)+\gamma=0, \ \forall  \pi_a\in \Delta\mathcal{B}.$$ Take $\pi^{ZD}(\eta,\beta,\gamma)$ as the corresponding ZD strategy. 
With the help of the ZD strategy's unilateral enforcement in players' utilities, we aim to investigate whether the defender can adopt the ZD strategy to better maintain its defensive performance than the original SSE strategy against a boundedly rational attacker.

%\section{Performance of ZD Strategy}

In what follows, we investigate the existence of ZD strategies to guarantee the availability for the defender.
\subsection{Existence of ZD Strategy}

Actually, the ZD strategy cannot enforce an arbitrary linear relation between two players’ utilities since it must belong to the implementer's strategy set. Thus, a feasible linear relation enforced by ZD strategies is fundamental for further analysis. The following lemma provides a necessary and sufficient condition for the feasibility of a linear relationship between players’ utilities, whose proof can
be found in Appendix \ref{app:L1}.

\begin{lemma}\label{th::ZD-1}
Under Assumption \ref{as::1}, there exists a ZD strategy which enforces  $\eta U_d+\beta U_a+\gamma=0$ 
if and only if  either of the following two inequalities is satisfied. \begin{equation}\label{eq::ZD-parameter}
   \max\limits_{s\in\{11,12\}} \eta U_{s}^d +\beta U_{s}^a \leqslant -\gamma\leqslant \min\limits_{s\in\{21,22\}} \eta U_{s}^d+\beta U_{s}^a.
\end{equation}
\begin{equation}\label{eq::ZD-parameter2}
\max\limits_{s\in\{21,22\}} \eta U_{s}^d+\beta U_{s}^a \leqslant-\gamma \leqslant   \min\limits_{s\in\{11,12\}} \eta U_{s}^d +\beta U_{s}^a.
\end{equation}
%with at least one inequality in (\ref{eq::ZD-parameter}) or (\ref{eq::ZD-parameter2}) being strict, respectively.
\end{lemma}

Lemma \ref{th::ZD-1} implies that the defender can adopt the ZD strategy $\pi^{ZD}(\eta,\beta,\gamma)$,  where $\eta$, $\beta$, and $\gamma$ satisfy (\ref{eq::ZD-parameter}) or (\ref{eq::ZD-parameter2}), to enforce an ideal linear relation between players' excepted utilities.  Actually, Lemma \ref{th::ZD-1} extends the application of ZD strategies since the payoff matrix in security games is not as symmetric as that in IPD games  \cite{hilbe2015partners}. Moreover,  Lemma \ref{th::ZD-1} covers the following cases.

\begin{itemize}

\item The defender can unilaterally restrict attacker's utility if $U_{11}^a>U_{12}^a$.  If the defender takes $\pi^{ZD}(\eta,\beta,\gamma)$ with $\eta=0, \beta\neq 0$, and $   U_{12}^a \leqslant -\frac{\gamma}{\beta} \leqslant  U_{11}^a$, then the defender ZD  can unilaterally restrict the attacker's utility as $-\frac{\gamma}{\beta}\in [U_{11}^a,U_{12}^a]$, which is the same as the equalizer strategy in  IPD games \cite{press2012iterated,cheng2022misperception}. 

\item The defender can unilaterally restrict its own utility if $U_{12}^d>U_{22}^d$. If the defender takes  $\pi^{ZD}(\eta,\beta,\gamma)$ with $\eta\neq0, \beta=0$, and $ U_{22}^d \leqslant -\frac{\gamma}{\eta}\leqslant U_{12}^d$, then the defender can unilaterally restrict the defender's utility between $U_{22}^d$ and $U_{12}^d$, which is consistent with the result in MTD problems \cite{wang2019moving}. % It is also a valuable extension of ZD strategies since a player cannot unilaterally set its own utility by adopting ZD strategies in IPD games \cite{press2012iterated}.
\end{itemize}

Based on the existence condition of feasible linear relations enforced by ZD strategies, we can further analyze whether there exists at least one ZD strategy in the security game $\mathcal{G}$. For simplification, for any $x_1,x_2,y_1,y_2\in\mathbb{R}$ with $x_1\neq x_2$, denote 
$$\begin{aligned}\Gamma^{-}(x_1,y_1,x_2,y_2)=\{(x,y)|y-y_1\leqslant \frac{y_2-y_1}{x_2-x_1} (x-x_1)\},\\
\Gamma^{+}(x_1,y_1,x_2,y_2)=\{(x,y)|y-y_1\geqslant \frac{y_2-y_1}{x_2-x_1} (x-x_1)\}.\end{aligned}$$Actually, $\Gamma^{-}$ ($\Gamma^{+}$) is the region above (below) the line going through points $(x_1,y_1)$ and $(x_2,y_2)$.  Then the following theorem shows a sufficient condition for the existence of  a ZD strategy in $\mathcal{G}$, whose proof can be found in Appendix \ref{app::T1}.

\begin{theorem}\label{th:ZD-2}
Under Assumption \ref{as::1}, there exists at least one ZD strategy of the defender in $\mathcal{G}$ if 
either of the following two relations is satisfied. 
\begin{align*}(U_{21}^d,U_{21}^a) ,(U_{22}^d,U_{22}^a) \in \Gamma^{+}(U_{11}^d,U_{11}^a,U_{12}^d,U_{12}^a).\\
(U_{21}^d,U_{21}^a) ,(U_{22}^d,U_{22}^a) \in \Gamma^{-}(U_{11}^d,U_{11}^a,U_{12}^d,U_{12}^a).
\end{align*}
%$$(U_{21}^d,U_{21}^a) ,(U_{22}^d,U_{22}^a) \in \Gamma^{-}(U_{11}^d,U_{11}^a,U_{12}^d,U_{12}^a).$$
\end{theorem}

\begin{figure}[tbp]
\centering
\includegraphics[width=2.8in]{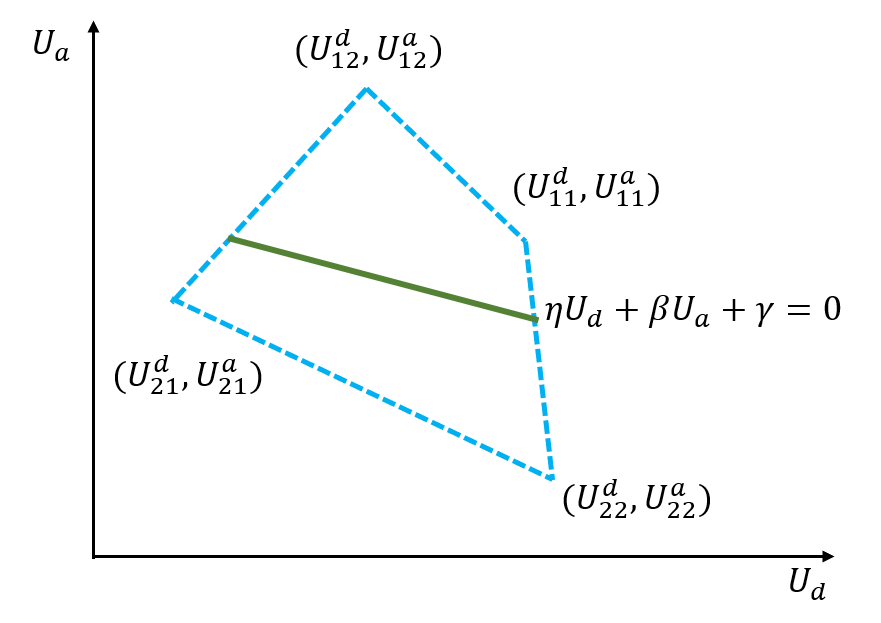}

\caption{ $(U_{11}^d,U_{11}^a)$ and $(U_{12}^d,U_{12}^a)$ lie in the one side of the line $\eta U_a +\beta U_b+\gamma=0$, while $(U_{21}^d,U_{21}^a)$ and $(U_{22}^d,U_{22}^a)$ lie in the other side. 
} 
\label{fi::A1}
\end{figure}

Theorem \ref{th:ZD-2} shows that, in the security game $\mathcal{G}$, the defender is able to choose a ZD strategy once $(U_{21}^d,U_{21}^a) ,(U_{22}^d,U_{22}^a)$ lie in the same side of the line going through points $(U_{11}^d,U_{11}^a)$ and $(U_{12}^d,U_{12}^a)$, as shown in Fig \ref{fi::A1}. %Otherwise, no ZD strategy is feasible for the defender and it cannot enforce a linear relation between players' expected utilities by adopting ZD strategies.
Hence, in this paper, we focus on the situation where there exists at least a ZD strategy, since selecting ZD strategies for the defender is based on its existence. In fact,  the two conditions are interchangeable. For example, if $(U_{21}^d,U_{21}^a) ,(U_{22}^d,U_{22}^a) \in \Gamma^{-}(U_{11}^d,U_{11}^a,U_{12}^d,U_{12}^a)$,  then can get $(U_{21}^d,U_{21}^a),(U_{22}^d,U_{22}^a) \in \Gamma^{-}(U_{11}^d,U_{11}^a,U_{12}^d,U_{12}^a)$ by swaping the values of $(U_{11}^d,U_{11}^a)$ and $(U_{22}^d,U_{22}^a)$, and swaping the values of $(U_{12}^d,U_{12}^a)$ and $(U_{21}^d,U_{21}^a)$. Hence, without loss of generality, the rest results in this paper are established with the condition  $(U_{21}^d,U_{21}^a),(U_{22}^d,U_{22}^a) \in \Gamma^{-}(U_{11}^d,U_{11}^a,U_{12}^d,U_{12}^a)$.

\subsection{ZD Strategy in Two Special Cases}

For understanding easily, we start with two special cases: $\lambda=1$, where the attacker takes BR strategies \cite{li2019cooperation}, and $\lambda=0$, where the attacker is a stubborn player with stubborn strategies \cite{la2016deceptive,nayak2016stubborn}. 

\textbf{When} $\bm{\lambda=1}$, the attacker chooses $ \textbf{{BR}}(\pi_d)$ after observing the defender's strategy $\pi_d$.  Recall the definition of $(\pi_d^{SSE},\pi_a^{SSE})$ as a SSE  and $U_d^{SSE}$ as the defender's utility from an SSE strategy. %  The next lemma shows that the defender cannot get a higher utility by adopting ZD strategies than that by adopting  SSE strategies.

\begin{lemma}\label{le::ZDlemma2}
Under Assumption \ref{as::1},  for any $\pi_d^{ZD}\in \Xi$ and $\pi_a\in BR(\pi_d^{ZD})$, $U_d(\pi_d^{ZD},\pi_a)\leqslant U_d(\pi_d^{SSE},\pi_a^{SSE})$. 
\end{lemma}

Lemma \ref{le::ZDlemma2} reveals that the SSE strategy always brings the highest utility for the defender when facing an attacker with the BR strategy. In this case, the upper limit of the performance of ZD strategies cannot surpass the defender's utility with adopting an SSE strategy.
In spite of this, the following theorem tells that ZD strategies admit a bounded loss compared with SSE strategies,  whose proof can be found in Appendix \ref{app::T2}. % and help defender quickly make decisions. The infimum of the loss and the corresponding ZD strategies are shown in the following theorem.

\begin{figure*}[tbp]
\centering
\subfigure[Attacker with the BR strategy]{
\begin{minipage}[t]{0.33\linewidth}
\centering
\includegraphics[width=2in]{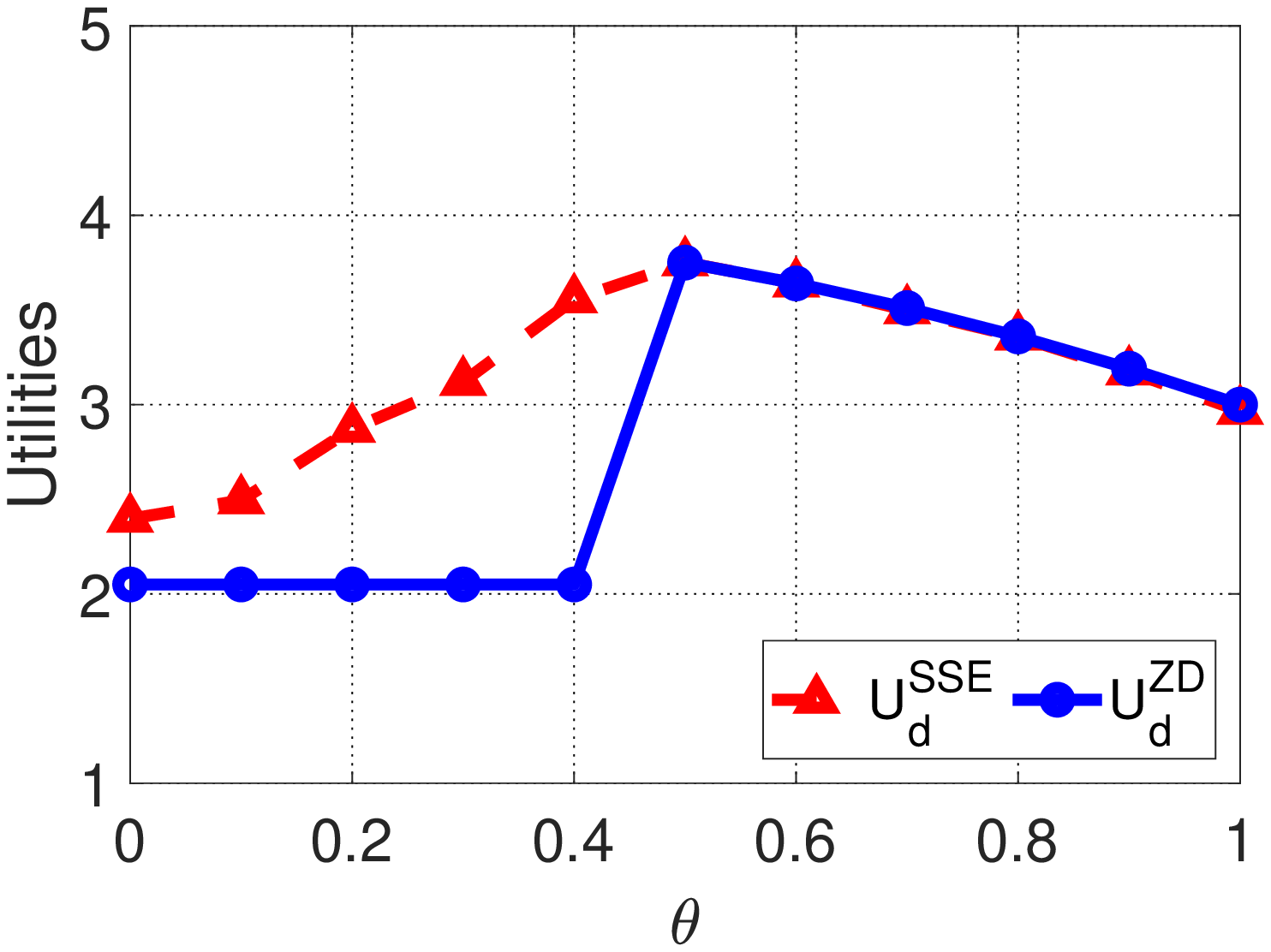}
%\caption{fig1}
\end{minipage}%
}%
\subfigure[Attacker with the stubborn strategy]{
\begin{minipage}[t]{0.33\linewidth}
\centering
\includegraphics[width=2in]{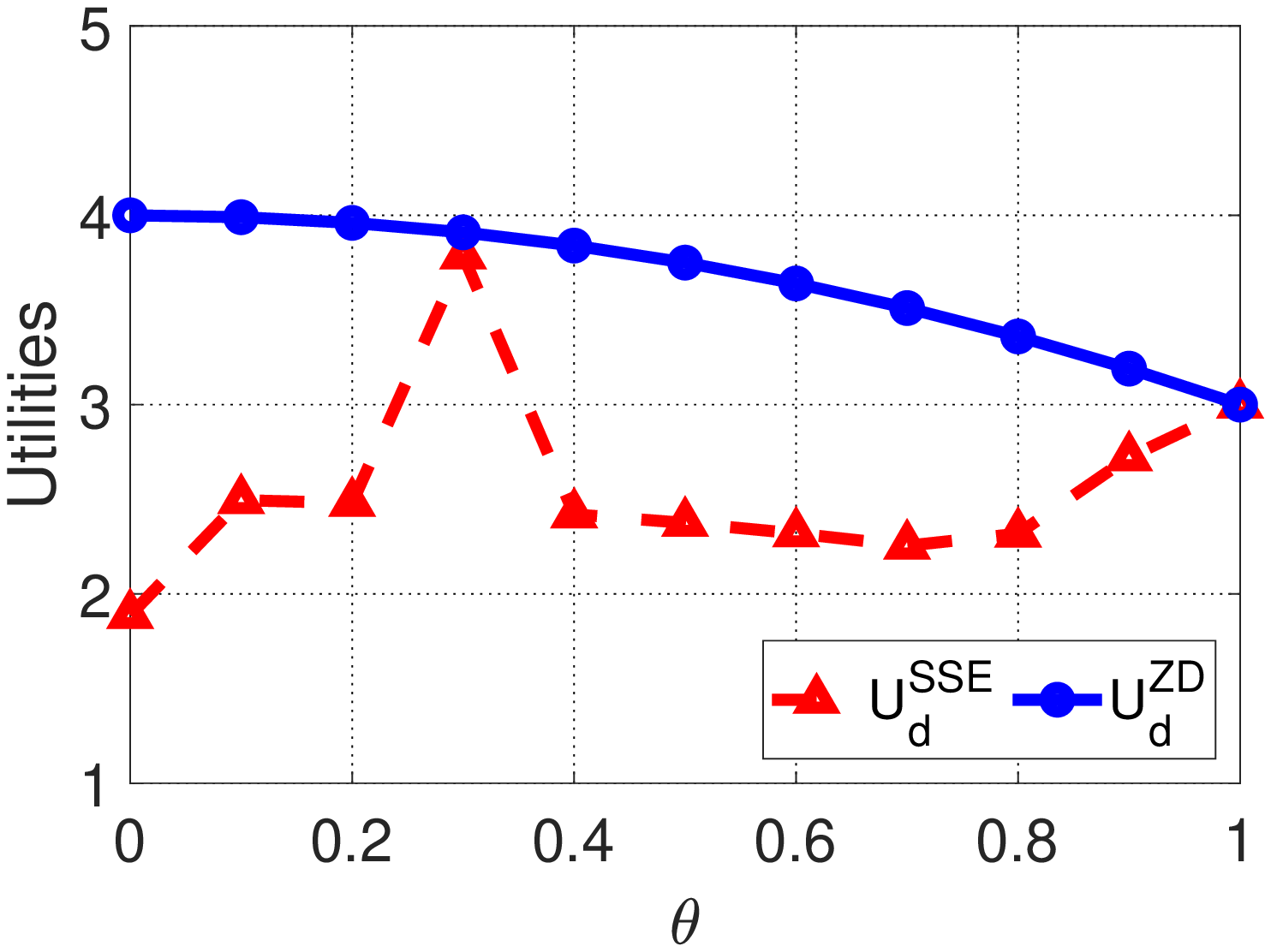}
%\caption{fig2}
\end{minipage}%
}%
\subfigure[Boundedly Rational Attacker]{
\begin{minipage}[t]{0.33\linewidth}
\centering
\includegraphics[width=2in]{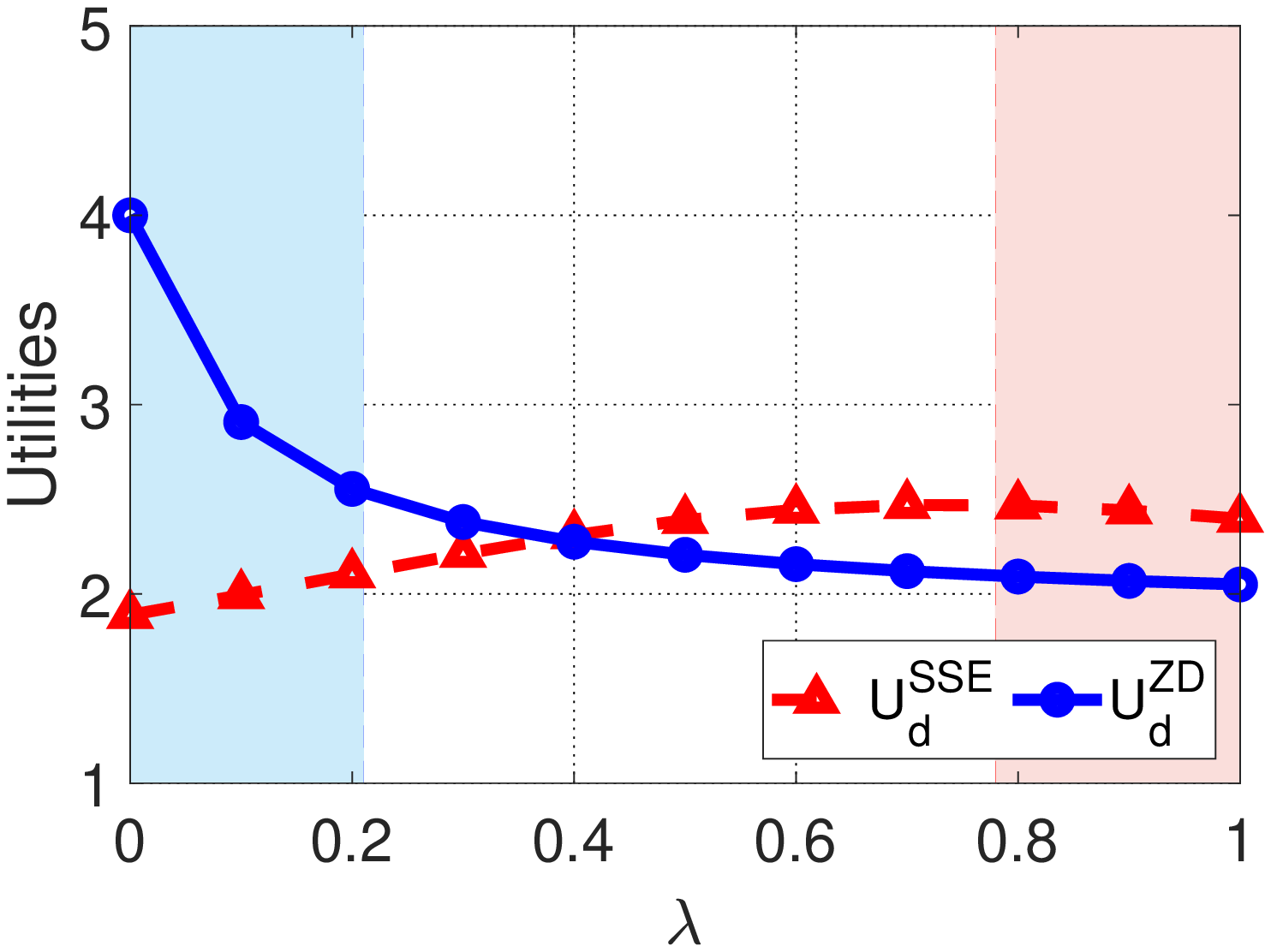}
%\caption{fig2}
\end{minipage}
}%
\centering
\caption{Performance of the ZD strategy compared with the SSE strategy in MTD problems. Red dotted lines describe the defender's expected utilities adopting an SSE strategy, while blue solid lines describe them adopting the corresponding ZD strategy. In (c), the red (blue) region shows the bound of $\Gamma_1$ ($\Gamma_2$) according to Theorem \ref{th::SSEbetter} (Theorem \ref{th::lambda}).}
\label{fi::MTD_problem}
\end{figure*}

\begin{theorem}\label{th::ZD-SSE-1}
Under Assumption \ref{as::1},
$$\begin{aligned}&\min\limits_{\pi_d^{ZD}\in \Xi}U_d\left(\pi_d^{SSE}, \pi_a^{BR}(\pi_d^{SSE})\right)\!-\!U_d\left(\pi_d^{ZD}, \pi_a^{BR}(\pi_d^{ZD})\right)\\
&= \left\{
\begin{array}{ll}
0, &\quad \quad\quad\text{if }  U_{11}^d\geqslant U_{21}^d,\\
U_d^{SSE}-U_{12}^d, &\quad \quad\quad \text{if }U_{11}^d < U_{21}^d.
\end{array}
\right.
\end{aligned}$$
The corresponding ZD strategy is $\pi_d^{ZD}$
$$=\! \!\left\{
\begin{array}{ll}
\pi^{ZD}\!(\!-\!k_1,\! 1,\! k_1U_{11}^d \!\!-\! U_{11}^a\!)
, &\!\! \text{if} \ U_{11}^d \!\geqslant\! U_{22}^d, \ U_{11}^a\!\geqslant\! U_{21}^a,\\
\pi^{ZD}(0,\!1,\!-U_{21}^a), &\!\!\text{if} \ U_{11}^d\!< \!U_{22}^d,\ U_{11}^a\!\geqslant\! U_{21}^a,\\
\pi^{ZD}\!(\!-\!k_2,\!1,\! k_2 U_{12}^d\!\!-\!U_{12}^a\!),  &\!\!\text{otherwise,}
\end{array}
\right.
$$ where $0\leqslant k_1\leqslant \frac{U_{11}^a-U_{21}^a}{U_{11}^d-U_{21}^d} $,  and $\frac{U_{22}^a-U_{12}^a}{U_{22}^d-U_{12}^d} \leqslant k_2\leqslant\frac{U_{11}^a-U_{12}^a}{U_{11}^d-U_{12}^d}$.

\end{theorem}

Theorem \ref{th::ZD-SSE-1} shows that the defender can adopt ZD strategies to get an tolerable loss in the utility compared with SSE strategies. On the one hand, when $U_{11}^a\geqslant U_{21}^a$, the ZD strategy in Theorem \ref{th::ZD-SSE-1} is an SSE strategy and brings the defender the same utility as SSE strategies. On the other hand, if the defender can endure the bounded loss,  then adopting the corresponding ZD strategy is also a good choice to avoid the complex calculation for SSE strategies, since deriving SSE strategies needs solve a bi-level optimization problem.

 \textbf{When} $\bm{\lambda=0}$, the attacker chooses  the stubborn strategy $\pi_a^*$.  The SSE strategy may not bring the defender a tolerable utility, since the stubborn attacker does not choose the BR strategy as the defender's expectation. In this case, due to the ZD strategy’s unilateral enforcement in players’ utilities, a ZD strategy  can exactly play an essential role to enforce desired utilities for the defender and even to bring a higher utility than the original SSE strategy does, and the following theorem's proof can be found in Appendix \ref{app::T3}

\begin{theorem}\label{th::stubborn}
Under Assumptions \ref{as::1} and \ref{as::pi}, there exists a ZD strategy $\pi_d^{ZD}= \pi^{ZD}(-k, 1, kU_{12}^d -U_{12}^a)$ with $k=\frac{U_{11}^a-U_{12}^a}{U_{11}^d-U_{12}^d}$ such that $$U_d(\pi_d^{ZD},\pi_a^*)\geqslant U_d(\pi_d^{SSE},\pi_a^*).$$

\end{theorem}

%Theorem \ref{th::stubborn} shows that, when facing a stubborn attacker, the defender can adopts ZD strategies to get a higher utility than adopting SSE strategies, which is contrary to Theorem \ref{th::ZD-SSE-1} when facing a learning attacker. 

In fact, the ZD strategy $\pi^{ZD}(-k, 1, kU_{12}^d -U_{12}^a)$ enforces the linear relation between players' expected utilities going through $(U_{12}^a,U_{12}^b)$ and $(U_{11}^d,U_{11}^a)$. The infimum of the ZD strategy $\pi^{ZD}(-k, 1, kU_{12}^d -U_{12}^a)$ is not lower than that of any other ZD strategy, including the ones which can unilaterally set the defender's utility \cite{wang2019moving}.  Moreover, according to its proof, when facing the stubborn attacker, this ZD strategy $\pi_d^{ZD}$ brings an increase $U_{11}^d-\frac{U_{11}^d \pi_d^{SSE}(1|21)+U_{21}^d\pi_d^{SSE}(2|11)}{\pi_d^{SSE}(2|11)+\pi_d^{SSE}(1|21)}$ in utility for the defender compared with the SSE strategy.

\subsection{ZD Strategy in General Case}

It is time to consider the general case \textbf{when} $\bm{\lambda\in[0,1]}$. Here, the boundedly rational attacker chooses the BR strategy with probability $\lambda$ and the stubborn strategy $\pi_a^*$ with probability $1-\lambda$, i.e.,  $\pi_a^{\lambda}(\pi_d,\pi_a^*)$ in Definition \ref{def::2}. Intuitively, the ZD strategy may bring the defender a similar performance as shown in Theorem \ref{th::ZD-SSE-1}  when $\lambda$ is close to $1$, and a similar performence as given in Theorem \ref{th::stubborn} when $\lambda$ is close to $0$. In fact, one main result of this subsection is given in the following theorem,  whose proof can be found in Appendix \ref{app::T4}.

%As discussed before, the ZD strategy is not worse than the SSE strategy for the defender if $\lambda=1$, while the SSE strategy is not worse than the ZD strategy for the defender if $\lambda=0$. Similar to Theorem \ref{th::ZD-SSE-1}, even though ZD strategies may not be better than SSE strategies, ZD strategies bring the defender's utility a bounded loss in defensive performance with SSE strategies, and can help the defender quickly make decisions, as shown in the following theorem.

\begin{figure*}[tbp]
\centering
\subfigure[Fictitious play with $\lambda=0.1$]{
\begin{minipage}[t]{0.24\linewidth}
\centering
\includegraphics[width=1.85in]{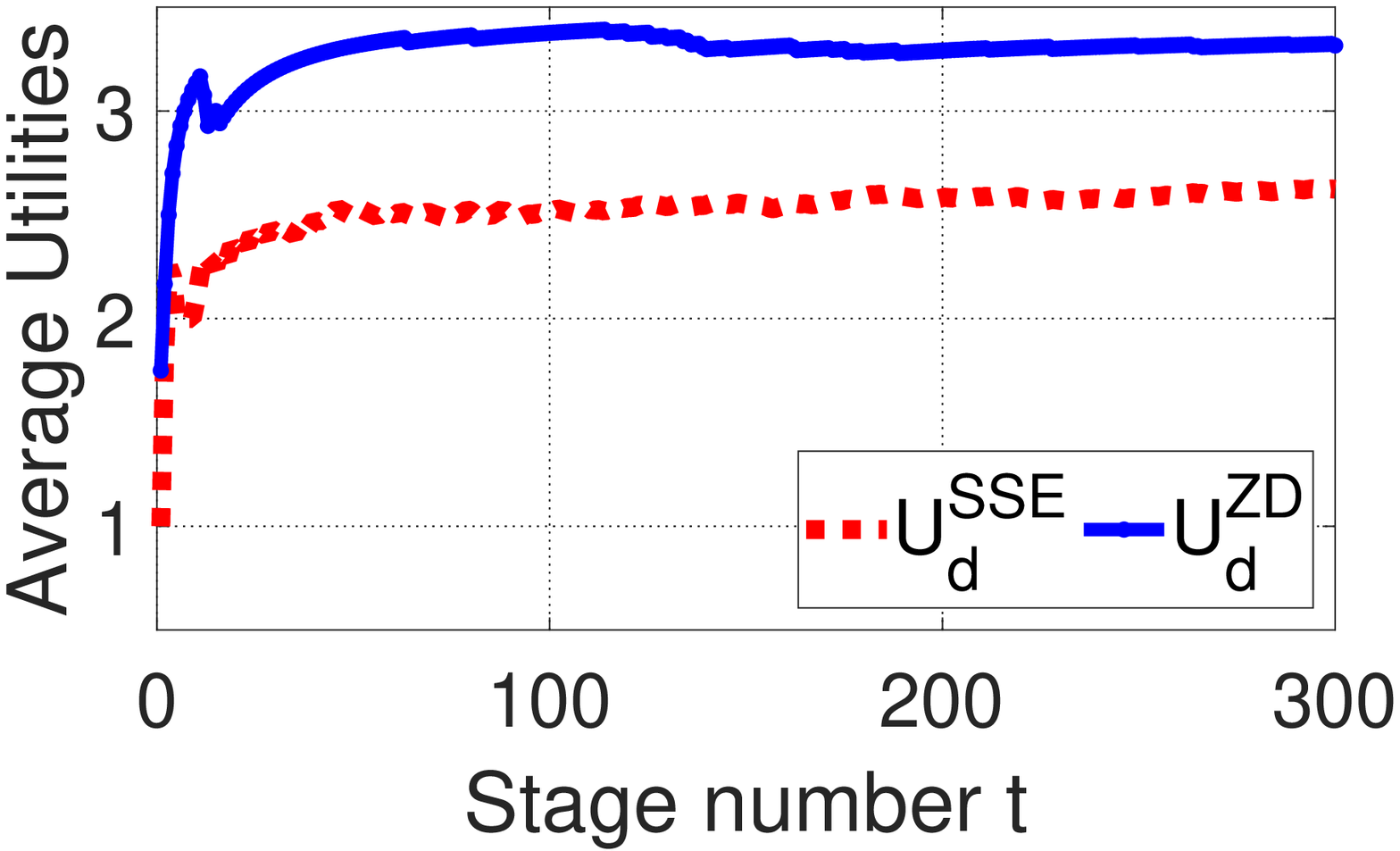}
%\caption{fig1}
\end{minipage}%
}%
\subfigure[Fictitious play with $\lambda=0.2$]{
\begin{minipage}[t]{0.24\linewidth}
\centering
\includegraphics[width=1.85in]{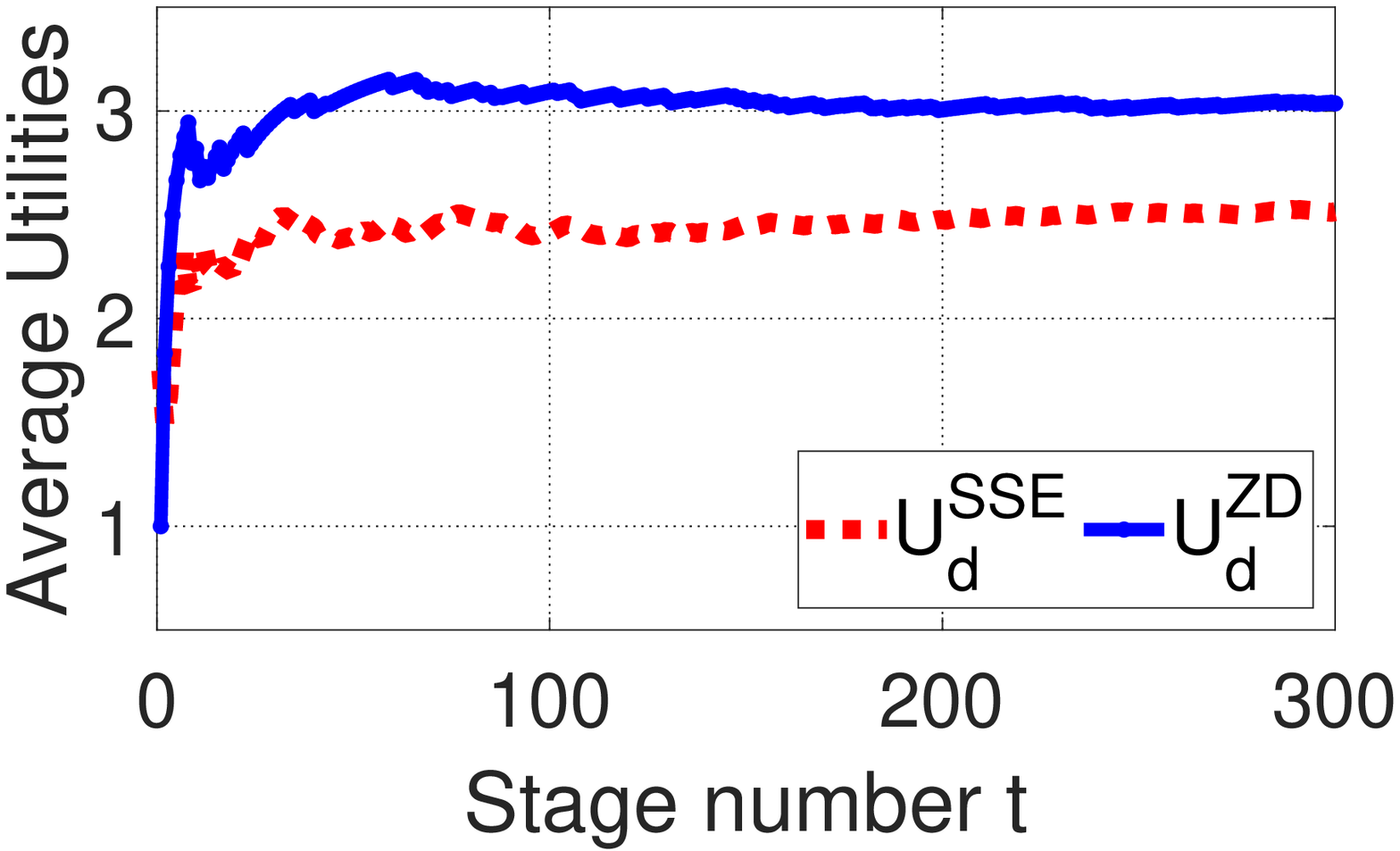}
%\caption{fig2}
\end{minipage}
}%
\subfigure[Fictitious play with $\lambda=0.8$]{
\begin{minipage}[t]{0.24\linewidth}
\centering
\includegraphics[width=1.85in]{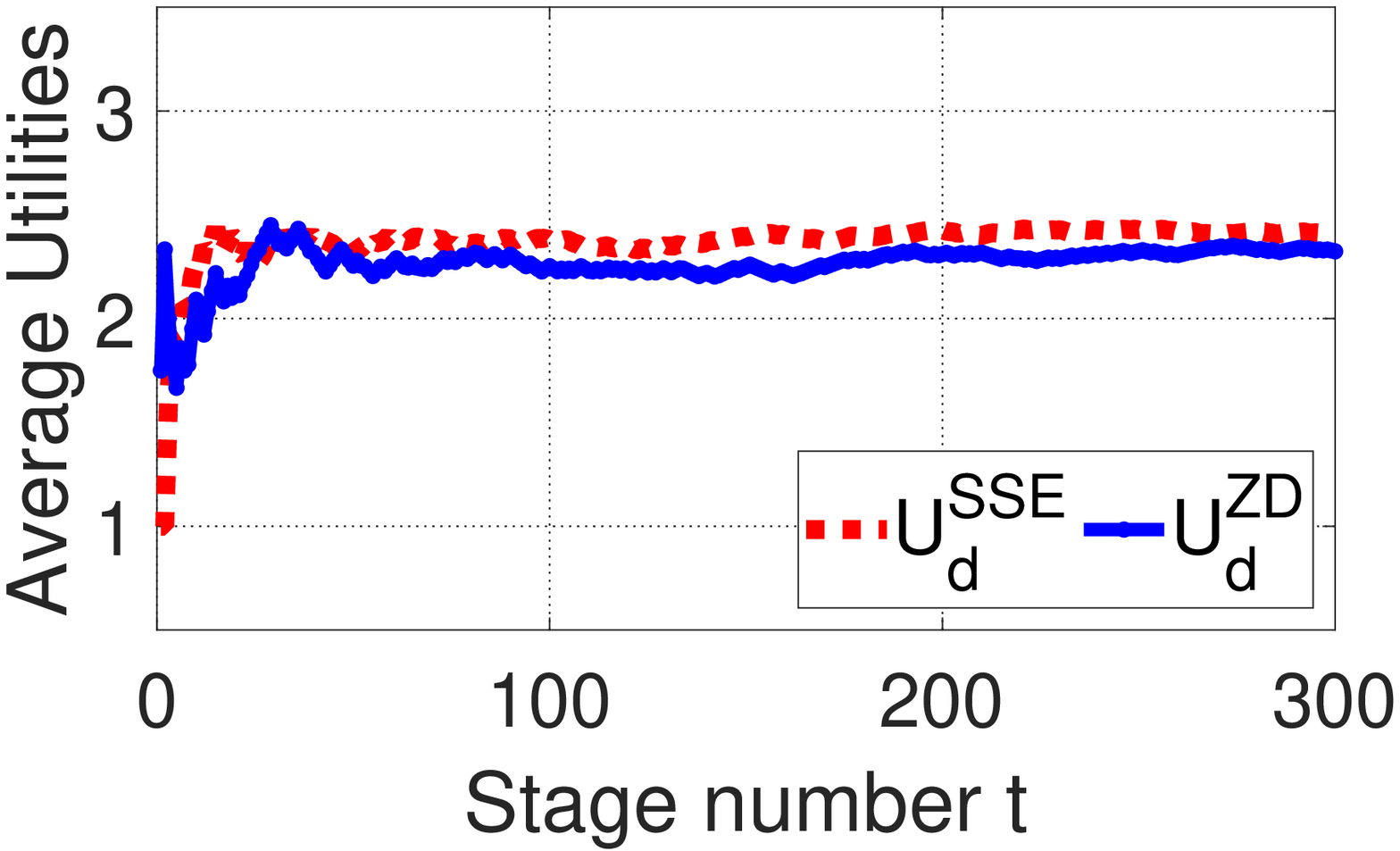}
%\caption{fig1}
\end{minipage}%
}%
\subfigure[Fictitious play with $\lambda=0.9$]{
\begin{minipage}[t]{0.24\linewidth}
\centering
\includegraphics[width=1.85in]{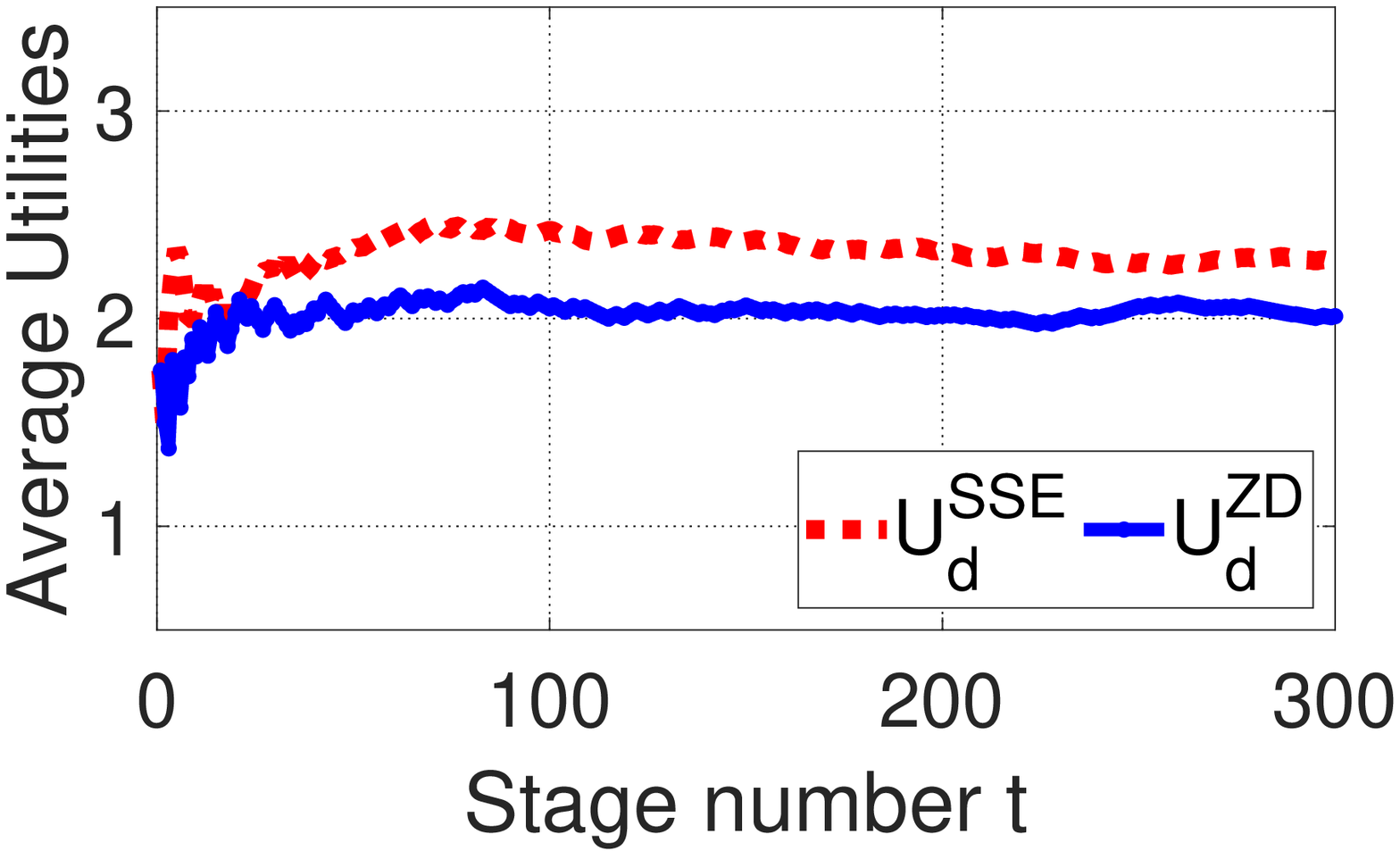}
%\caption{fig2}
\end{minipage}
}
\subfigure[Q-learning with $\lambda=0.1$]{
\begin{minipage}[t]{0.24\linewidth}
\centering
\includegraphics[width=1.85in]{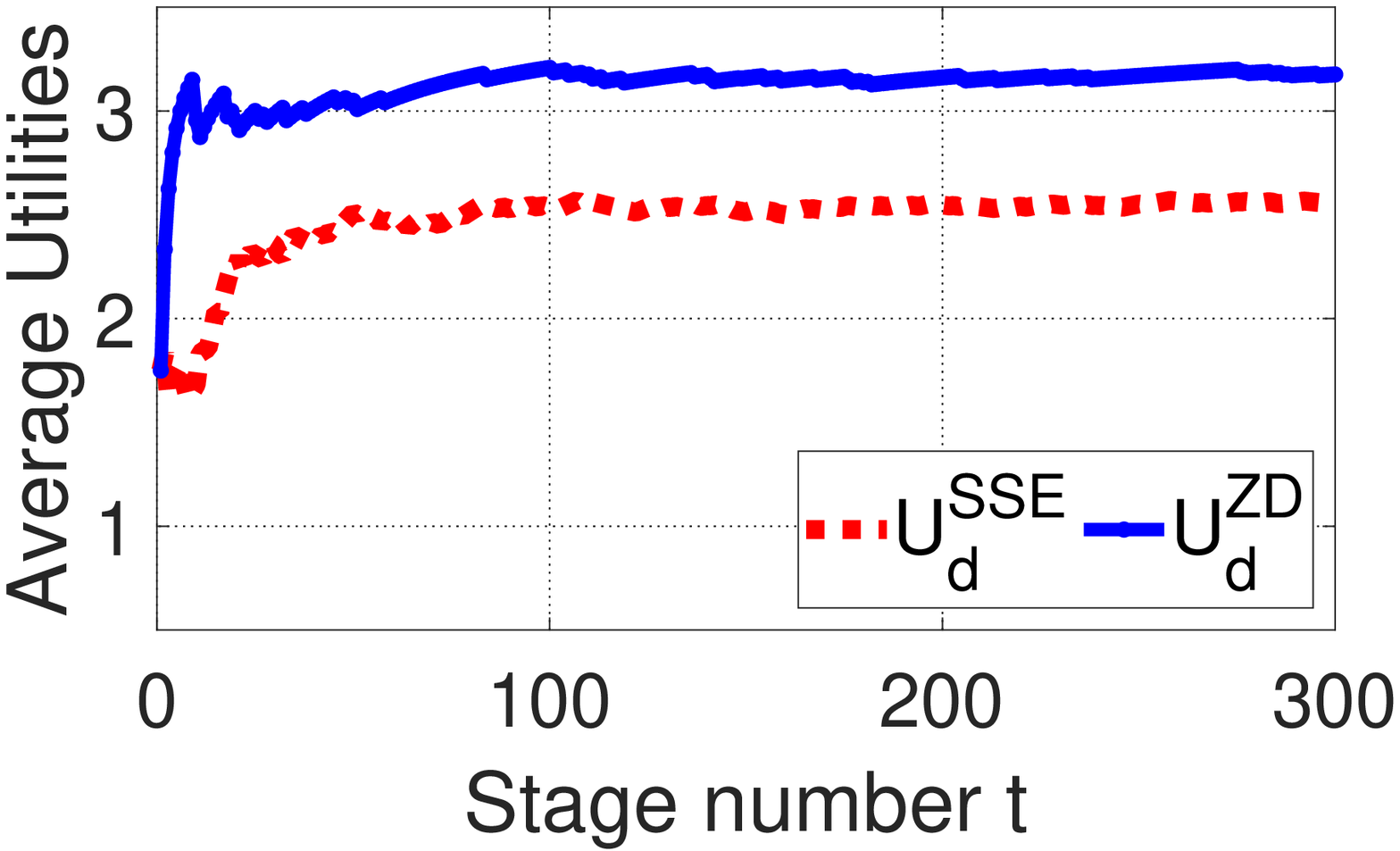}
%\caption{fig1}
\end{minipage}%
}%
\subfigure[Q-learning with $\lambda=0.2$]{
\begin{minipage}[t]{0.24\linewidth}
\centering
\includegraphics[width=1.85in]{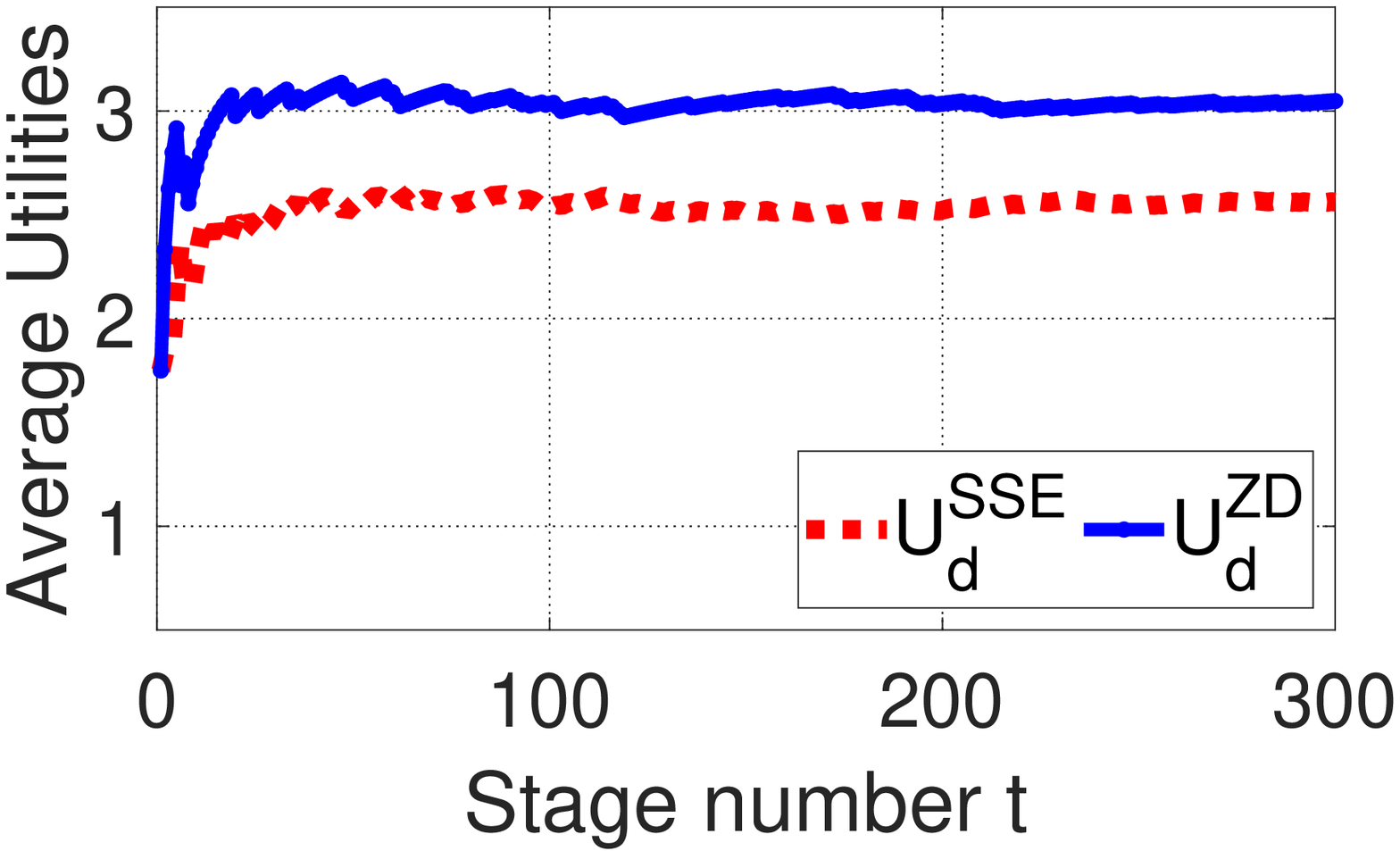}
%\caption{fig2}
\end{minipage}
}%
\subfigure[Q-learning with $\lambda=0.8$]{
\begin{minipage}[t]{0.24\linewidth}
\centering
\includegraphics[width=1.85in]{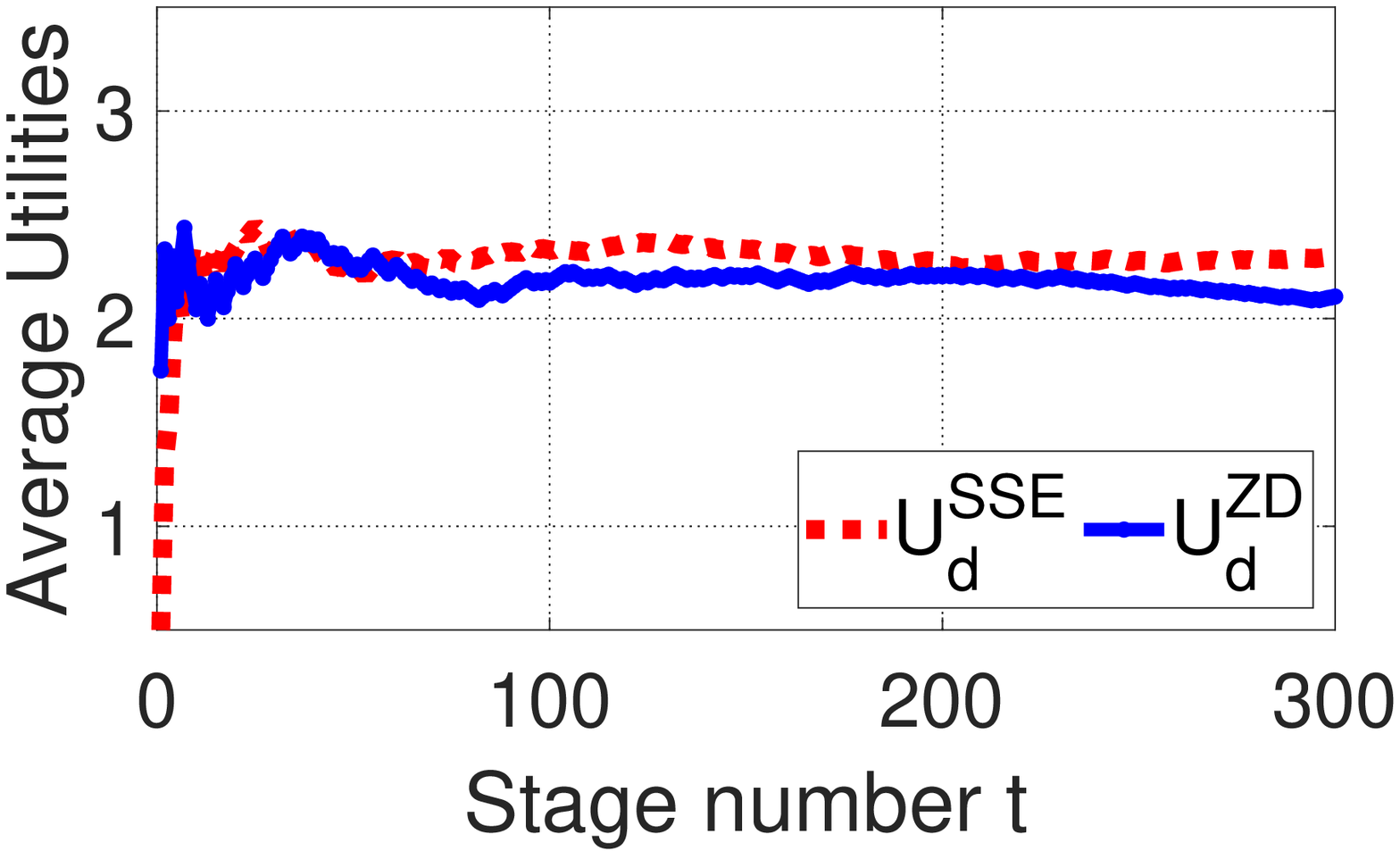}
%\caption{fig1}
\end{minipage}%
}%
\subfigure[Q-learning with $\lambda=0.9$]{
\begin{minipage}[t]{0.24\linewidth}
\centering
\includegraphics[width=1.85in]{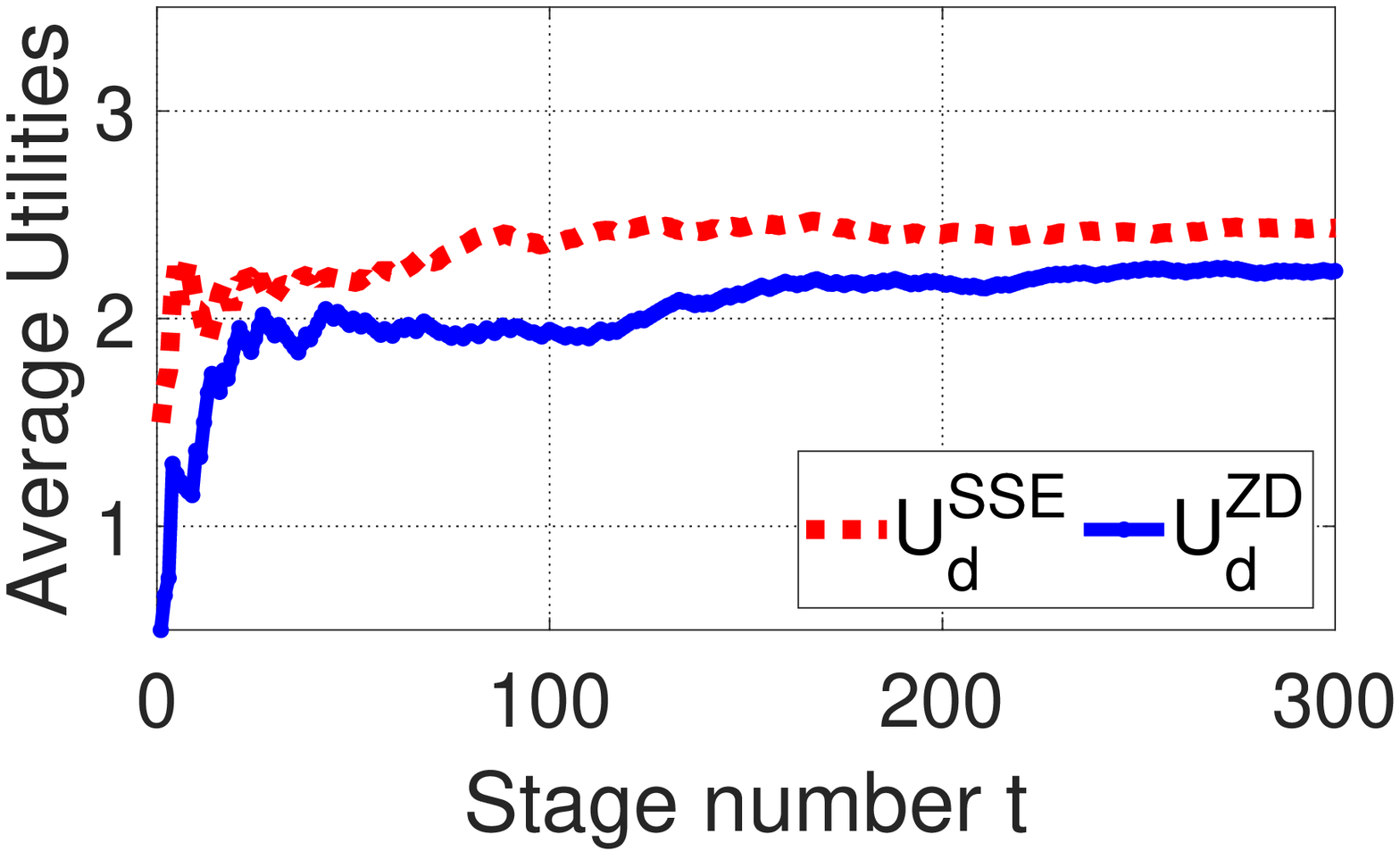}
%\caption{fig2}
\end{minipage}
}
\centering
\caption{Performance of the ZD strategy compared with the SSE strategy in CPS with different mechanisms. Red dotted lines show the defender's average utility adopting an SSE strategy, while blue solid lines show the defender's average utility adopting the corresponding ZD strategy in Theorem \ref{th::SSEbetter} or Theorem \ref{th::lambda}.}
\label{fi::Learning}
\end{figure*}

\begin{theorem}\label{th::SSEbetter}
Under Assumptions \ref{as::1} and \ref{as::pi}, %if $\lambda\in\Gamma_1$, then for any  $\pi_d^{ZD}\in \Xi$, $$U_d(\pi_d^{ZD},\pi_a^{\lambda}(\pi_d^{ZD},\pi_a^*))\!\leqslant \!U_d(\pi_d^{SSE},\pi_a^{\lambda}(\pi_d^{SSE},\pi_a^*)),$$ where $\Gamma_1$ is defined in (\ref{eq::Gamma1}), and the parameters therein are shown in the supplementary material. Moreover, 
for $\lambda\in \Gamma_1$ in (\ref{eq::Gamma1}),  there is
$$\begin{aligned}&\min\limits_{\pi_d^{ZD}\in \Xi}\! U_d\left(\pi_d^{SSE}\!,\!\pi_a^{\lambda}(\pi_d^{SSE}\!,\!\pi_a^*)\!\right)\!- \!U_d\left(\pi_d^{ZD}\!,\!\pi_a^{\lambda}(\pi_d^{ZD},\!\pi_a^*)\!\right)\\
&\leqslant\!\!\left\{
\begin{array}{ll}
0, & \text{if} \ U_{11}^d\!\geqslant\! U_{22}^d, \ U_{11}^a\!\geqslant \!U_{21}^a,  \\
H(\pi_d^{ZD},\pi_d^{SSE},\pi_a^*,\lambda), & \text{otherwise},
\end{array}
\right.
\end{aligned}$$
where $H(\pi_d^{ZD},\pi_d^{SSE},\pi_a^*,\lambda)$ was defined in (\ref{eq::H}), and the parameters therein are shown in  Appendix \ref{app::No}. The corresponding ZD strategy is $\pi_d^{ZD}$
$$ =\!\!\left\{\!
\begin{array}{ll}
\!\pi^{ZD}(-k_1,\! 1,\! k_1U_{11}^d \!-\!U_{11}^a)
, & \text{if} \ U_{11}^d\! \geqslant \!U_{22}^d, \ U_{11}^a\!\geqslant \!U_{21}^a,\\
%\!\pi^{ZD}(0,\!1,\!-\!U_{21}^a), &\!\!\!\!\!\!\!\!\!\!\!\!\!\!\!\!\!\!\!\!\text{if}\ \lambda\!=\!1, U_{11}^d\!<\! U_{22}^d, \text{and} U_{11}^a\!\geqslant \!U_{21}^a\!,\\
\!\pi^{ZD}(-k_2,\!1, \!k_2 U_{12}^d\!-\!U_{12}^a), &\text{otherwise},
\end{array}
\right.
$$ where $0\leqslant k_1\leqslant \frac{U_{11}^a-U_{21}^a}{U_{11}^d-U_{21}^d} $, and $k_2=\frac{U_{11}^a-U_{12}^a}{U_{11}^d-U_{12}^d}$.

\end{theorem}

Theorem \ref{th::SSEbetter} provides the set $\Gamma_1$ for the defender, in which the ZD strategy brings a bounded and tolerable loss in defensive performance, even though the ZD strategy cannot surplus the SSE strategy. Thus, for the boundedly rational attacker with $\lambda\in \Gamma_1$, if the defender does not care too much about losing a little utility, then the defender can adopt the corresponding ZD strategy since adopting the ZD strategy in Theorem \ref{th::SSEbetter}  avoids paying vast resources to solve a bi-level optimization problem for the SSE strategy. Moreover, if $U_{11}^a\geqslant U_{21}^a$ and $U_{11}^d\geqslant U_{22}^d$, the ZD strategy can bring the defender the same utility as an SSE strategy, which means that the defender can still adopt ZD strategies.

 Although $\Gamma_1$ seems complicated to verify, some typical value of $\lambda$ is easy to be confirmed whether it belongs to $\Gamma_1$. For instance, $\lambda=1$ is always in $\Gamma_1$. In this case, the attacker tends to take the BR strategy,  which is consistent with Theorem \ref{th::ZD-SSE-1}. Actually, $\lambda$ is in $\Gamma_1$ if $\lambda$  is close to $1$, which means that the attacker tends to choose the BR strategy. Also, we provide a subset of $\Gamma_1$, which can be verified easily by the defender.

\begin{corollary}

Under Assumptions \ref{as::1} and \ref{as::pi}, if $\lambda\in[\frac{1}{2},1]$ and $$\begin{aligned}&4(U_{11}^d-U_{21}^d)\pi_{d}^{SSE}(2|11)(1-\lambda)^4\\
\leqslant\!&\! \left(\frac{1}{4}(U_d^{SSE}\!\!-\!U_{12}^d)D(\textbf{1})\!-\! B \right)\!\!\left(\pi_d^{SSE}(2|11)\!+\!\pi_d^{SSE}(1|21)\right), \end{aligned}$$
then
$$\begin{aligned}&\min\limits_{\pi_d^{ZD}\in \Xi}\! U_d\left(\pi_d^{SSE}\!,\!\pi_a^{\lambda}(\pi_d^{SSE}\!,\!\pi_a^*)\!\right)\!- \!U_d\left(\pi_d^{ZD}\!,\!\pi_a^{\lambda}(\pi_d^{ZD},\!\pi_a^*)\!\right)\\
&\leqslant\!\!\left\{
\begin{array}{ll}
0, & \text{if} \ U_{11}^a\!\geqslant \!U_{21}^a, \ U_{11}^d\!\geqslant\! U_{22}^d,\\
H(\pi_d^{ZD},\pi_d^{SSE},\pi_a^*,\lambda), & \text{otherwise}.
\end{array}
\right.
\end{aligned}$$
                                                                
\end{corollary}

At last, we analogously consider the situation when $\lambda$ is close to $0$, i.e., the attacker tends to take the stubborn strategy in the following theorem,  whose proof can be found in Appendix \ref{app::T5}.

%On the other hand,  the ZD strategy may be better than SSE strategies for the defender.  Thus, it is significant for the defender to salfly to chooses ZD strategy for the boundedly rational attacker with special $\lambda$, as shown in the following theorem.

\begin{theorem}\label{th::lambda}
Under Assumptions \ref{as::1} and \ref{as::pi}, if $\lambda \in \Gamma_2 $ in (\ref{eq::Gamma1}), then there exists  a ZD strategy $\pi_d^{ZD}=\pi^{ZD}(-k, 1, k U_{21}^a -U_{21}^b)$ such that $$U_d(\pi_d^{ZD},\pi_a^{\lambda}(\pi_d^{ZD},\pi_a^*))\!\geqslant \!U_d(\pi_d^{SSE},\pi_a^{\lambda}(\pi_d^{SSE},\pi_a^*)),$$
where  $k=\frac{U_{11}^b-U_{12}^b}{U_{11}^a-U_{12}^a}.$%, and $\Gamma_2$ is defined in (\ref{eq::Gamma1}).
\end{theorem}

Theorem \ref{th::lambda} provides the set $\Gamma_2$ for the defender to adopt the ZD strategy to get a higher utility than the original SSE strategy. Notice that the corresponding ZD strategy yields wonderful performance. If $\lambda\in\Gamma_2$, the defender can confidently select the corresponding ZD strategy, since the ZD strategy brings the defender higher defensive performance than an SSE strategy does. 

%Notice that $\Gamma_1$ and $\Gamma_2$ have no common element when $(U_{12}^a-U_{SSE})D(\textbf{1})\neq 0$ and $A\neq 0$. 
 Clearly, $\lambda=0$ is always in $\Gamma_2$, which is consistent with the results  in Theorem \ref{th::stubborn}. Actually, $\lambda$ is in $\Gamma_2$ if $\lambda$  is close to $0$, which means that the attacker tends to be a stubborn attacker. Also, we provide a subset of $\Gamma_2$ for the defender, which can be verified easily by the defender.

\begin{corollary}

Under Assumptions \ref{as::1} and \ref{as::pi}, if $\lambda\in[0,\frac{1}{2}]$ and $$4(U_d^{SSE}-U_{12}^a)D(\textbf{1})\lambda^4\leqslant \frac{1}{4}A-B, $$
then there exists   $\pi_d^{ZD}\in \Xi$ such that $$U_d(\pi_d^{ZD},\pi_a^{\lambda}(\pi_d^{ZD},\pi_a^*))\!\geqslant \!U_d(\pi_d^{SSE},\pi_a^{\lambda}(\pi_d^{SSE},\pi_a^*)).$$
                                                                
\end{corollary}

\section{Applications}

For illustration, we provide experiments to verify that the ZD strategy can help the defender maintain its defensive performance against a boundedly rational attacker, where the baseline is the original SSE strategy. %Here, the attacker can directly observe the defender's strategy in MTD problems and only receive players’ action history in  CPS.

\subsection{In MTD Problems}

Let us consider an MTD problem, where the attacker can directly observe the defender's strategy and take its explicit BR strategy \cite{feng2017signaling,carvalho2014moving}. Take $Y_i$ as the cost of the defender moving the defense resource from target $i$ to the other target, and take $C_i$ as the cost of the attacker invading target $i$. Similar to \cite{wang2019moving}, we also use the average to approximate the transfer cost. %For instance, the average transfer cost in the state $11$ is $\frac{Y_2}{2}$. % since the defender may not only keep the defender resource but also transfer the resource from target $2$ to target $1$. 
Also, take $R^d_i$ and $R^a_i$ as the reward and loss for two players in the state $i\in\mathcal{S}$. Thus, the utility matrix in MTD problems is shown in Table \ref{tab::MTD}. Take $R_s^d=d_s^1\theta+d_s^0$, and $R_s^a=a_s^1\theta+a_s^0$ for any $s\in \mathcal{S}$, where $\theta\in[0,1]$, and  $d_s^k$ and $a_s^k$ are parameters in players' rewards and losses, respectively, for any $s\in \mathcal{S}$ and $k\in \{1,2\}$.
\begin{table}[h]
\renewcommand\arraystretch{1.3}
\tabcolsep=0.2cm
\caption{Utility matrix in MTD problems}  
\centering  

       \begin{tabular}{cccc}
\hline
& &\multicolumn{2}{c}{Attacker} \\
& &1&2 \\
\hline 
\multirow{2}*{Defender\!}& 1 & $(R^d_{11}\!-\!\frac{Y_2}{2},\!R^a_{11}\!-\!C_1)$ & $(R^d_{12}\!-\!\frac{Y_2}{2},R^a_{12}\!-\!C_2)$ \\
~&2 & $(R^d_{21}\!-\!\frac{Y_1}{2},\!R^a_{21}\!-\!C_1)$ & $(R^d_{22}\!-\!\frac{Y_1}{2},\!R^a_{22}\!-\!C_2)$ \\
\hline
\end{tabular} 
       \label{tab::MTD} 
\end{table}

As shown in Fig \ref{fi::MTD_problem}(a), for an attacker with the BR strategy, the expected utility of the defender with an SSE strategy is always higher than that with a ZD strategy. Besides, in Fig \ref{fi::MTD_problem}(b), for an attacker with the stubborn strategy, the expected utility of the defender with adopting the ZD strategy in Theorem \ref{th::stubborn} is always higher than that with adopting an SSE strategy. Moreover, in Fig \ref{fi::MTD_problem}(c), $U_d^{ZD}$ is always higher than $U_d^{SSE}$ when $\lambda<0.21$, which is consistent with Theorem \ref{th::lambda}. Also, $U_d^{SSE}$ is always higher than $U_d^{ZD}$ when $\lambda>0.78$, which is consistent with Theorem \ref{th::SSEbetter}, and the difference between the two utilities is bounded and tolerable.

\subsection{In CPS Problems}

Here we consider a CPS problem with a defender as a system administrator and an attacker as a jammer or an eavesdropper. Here, different from the previous experiment, the attacker can only observe players' action history without  directly receiving the defender's strategy. The attacker adopts the BR strategy based on the fictitious play \cite{qiu2021provably} or the Q-learning method \cite{li2019cooperation}, whose details are provided in Appendix \ref{app::Al}.

As shown in Fig \ref{fi::Learning}(a)-(f), when $\lambda$ is close to $0$, like $0.1$ in Fig \ref{fi::Learning}(a), (e),  and $0.2$ in Fig \ref{fi::Learning}(b), (f),  the average utility of the defender with the ZD strategy is higher than that with the SSE strategy.  Besides, when $\lambda$ is close to $1$, like $0.8$ in Fig \ref{fi::Learning}(c), (g),  and $0.9$ in Fig \ref{fi::Learning}(d), (h), although the average utility with the ZD strategy is lower than that with the SSE strategy, the loss is small enough to tolerate. Thus, the defender can also adopt ZD strategies to maintain its defensive performance with a bounded and tolerable loss, and avoid the complex computing in SSE strategies.
% \newline

\section{Discussion}

We have focused on stochastic Stackelberg asymmetric security games in this paper. Due to the stubborn elements within the boundedly rational attacker, we have investigated the defensive performance of ZD strategies, and have analyzed whether the ZD strategies can make up for the deficiencies of SSE strategies in such circumstances. Also, we have provided experiments to support our methodology by employing proper ZD strategies for the defender.

%A security game with multiple targets is more practical and realistic than the two-target security game in this paper.
Actually, our results can be extended to some security problems with multiple targets. For example, consider the case that the targets in the security game can be divided into two categories. Each player's utilities are the same when choosing any two targets belonging to one category, while the player's utilities are different when choosing any two targets belonging to different categories. In such a situation, the defender can still adopt similar ZD strategies in this paper to improve its defensive performance.

Indeed, in general multi-target asymmetric stochastic security games, there are  challenges in applying the ZD strategy against boundedly rational attackers. We show the barrier from a simple viewpoint. With multi-target settings, the expected utility of the defender will be composed of complex polynomials in $\lambda$, and each polynomial is the determinant of a $(n^2 \times n^2)$ matrix with a very high degree. Applying ZD strategies in general multi-target security games is still an open problem and  deserves more and more exploration.
%Even though, we believe the contributions in this paper may provide some foundation for further study, and can also be applied to some special security problems with multiple targets. For example, the defender may also adopt ZD strategies to improve then defensive performance  when the targets in the security game can be divided into two categories and each player's utility is similar when it chooses different targets in the same category.

\begin{acknowledgments}
This work was supported by the National Key Research and Development Program of China under No 2022YFA1004700, the National Natural Science Foundation of China under No. 62173250, and Shanghai
Municipal Science and Technology Major Project under No. 2021SHZDZX0100.
\end{acknowledgments}
\begin{widetext}
\appendix

\section{\label{app::No}Notations}
 For any $\pi_d\in\Delta \mathcal{D},\pi_a^1,\pi_a^2\in\Delta \mathcal{A}$, and 
$\mathbf{f}=[f_1,f_2,f_3,f_4]^T\in\mathbb{R}^4$, denote
\begin{equation}\label{eq::D4}D(\pi_d,\pi_a^1,\pi_a^2,\mathbf{f})=\left[\begin{array}{llll}
\pi_d(1|11)\pi_a^2(1|11)-1  & \pi_d(1|11)-1 &\pi_a^1(1|11) -1 &f_1\\
\pi_d(1|12)\pi_a^2(1|12) & \pi_d(1|12)-1 & \pi_a^1(1|12)&f_2 \\
\pi_d(1|21)\pi_a^2(1|21) & \pi_d(1|21)&  \pi_a^1(1|21)-1&f_3\\
\pi_d(1|22)\pi_a^2(1|22) & \pi_d(1|22) &\pi_a^1(1|22) & f_4
\end{array}\right].\end{equation}
For convenience, we take $D(\pi_d,\pi_a,\mathbf{f})=D(\pi_d,\pi_a,\pi_a,\mathbf{f})$ and $D(\mathbf{f})=\max\limits_{\pi_d,\pi_a^1,\pi_a^2}D(\pi_d,\pi_a^1,\pi_a^2,\mathbf{f})$. 
Denote $$J(\pi_d,\pi_a,\mathbf{f})= D(\pi_d,\pi_a^{BR}(\pi_d),\pi_a,\mathbf{f})+D(\pi_d,\pi_a,\pi_a^{BR}(\pi_d),\mathbf{f}),$$ and $$C(\pi_d^{ZD},\pi_d^{SSE},\pi_a^*,\lambda)=D(\pi_d^{ZD},\lambda\pi_a^{BR}(\pi_d^{ZD})+(1-\lambda)\pi_a^*,\textbf{1})\cdot D(\pi_d^{SSE},\lambda\pi_a^{BR}(\pi_d^{SSE})+(1-\lambda)\pi_a^*,\textbf{1}),$$  to simplify the writing. Moreover, take $$
\begin{aligned}
A=&U_{11}^d-\frac{U_{11}^d \pi_d^{SSE}(1|21)+U_{21}^d\pi_d^{SSE}(2|11)}{\pi_d^{SSE}(2|11)+\pi_d^{SSE}(1|21)},\\
B_1=&\max\limits_{\pi_d^{SSE},\pi_d^{ZD}}\max\limits_{\mathbf{f}\in\{\textbf{1},S^d\}}\frac{1}{2}\left|D(\pi_d^{ZD},\pi_a^{BR}(\pi_d^{ZD}),S^d) J(\pi_d^{SSE},\pi_a^*,\mathbf{f})-  D(\pi_d^{SSE},\pi_a^{BR}(\pi_d^{SSE}),S^d) J(\pi_d^{ZD},\pi_a^*,\mathbf{f})\right|,\\
B_2=&\max\limits_{\pi_d^{SSE},\pi_d^{ZD}}\left| D(\pi_d^{ZD},\pi_a^*,S^d)  J(\pi_d^{SSE},\pi_a^*,\textbf{1}) + D(\pi_d^{SSE},\pi_a^*,\textbf{1}) J(\pi_d^{ZD},\pi_a^*,S^d)- D(\pi_d^{SSE},\pi_a^*,S^d)  J(\pi_d^{ZD},\pi_a^*,\textbf{1})\right. \\
& \left. \quad \quad\quad\quad  - D(\pi_d^{ZD},\pi_a^*,\textbf{1}) J(\pi_d^{SSE},\pi_a^*,S^d)\right|,\\
B_3=&\max_{\pi_d^{SSE},\pi_d^{ZD}}\left| J(\pi_d^{ZD},\pi_a^*,S^d) J(\pi_d^{SSE},\pi_a^*,\textbf{1})-  J(\pi_d^{ZD},\pi_a^*,S^d) J(\pi_d^{ZD},\pi_a^*,\textbf{1})\right|,\\
B=&\max\left\{B_1,B_2,\frac{1}{2}B_3\right\}.
\end{aligned}
$$

\section{\label{app:L1}Proof of Lemma 1}
Sufficiency:  Consider that the ZD strategy $\pi_d$ enforces $\eta U_d+\beta U_a+\gamma=0$.
Thus, $\pi_d(1)=\phi(\eta \mathbf{S}^{d}  +\beta \mathbf{S}^a +\gamma)+\hat{\pi}$, where $\phi\neq 0$. Since $\pi_d\in\Xi $, the following inequalities are satisfied:
\iffalse
$$
\begin{aligned}
-1\leqslant \phi(\eta U_{11}^a + \beta U_{11}^b + \gamma ) \leqslant 0\\
-1\leqslant \phi(\eta U_{12}^a + \beta U_{12}^b + \gamma ) \leqslant 0\\
0\leqslant \phi(\eta U_{21}^a + \beta U_{21}^b + \gamma ) \leqslant 1\\
0\leqslant \phi(\eta U_{22}^a + \beta U_{22}^b + \gamma ) \leqslant 1
\end{aligned}
$$
\fi
\begin{subequations}
\begin{equation}\label{eq::1a}-1\leqslant \phi(\eta U_{i}^d + \beta U_{i}^a + \gamma ) \leqslant 0, i\in\{11,12\},\end{equation}
\begin{equation}\label{eq::1d}0\leqslant \phi(\eta U_{j}^d + \beta U_{j}^a + \gamma ) \leqslant 1,j\in\{21,22\}.\end{equation}
\end{subequations}
If $\phi>0$, it follows from the right inequalities in (\ref{eq::1a}) and  the left inequalities in (\ref{eq::1d}) that 
\begin{equation}\label{eq::zhuanhua}
\begin{aligned}
\eta U_{i}^d + \beta U_{i}^a + \gamma  \leqslant 0, i\in\{11,12\},\\
\eta U_{j}^d + \beta U_{j}^a + \gamma  \geqslant 0,j\in\{21,22\},
\end{aligned}
\end{equation}
which implies $
\max (\eta U_{11}^d +\beta U_{11}^a, \eta U_{12}^d+\beta U_{12}^a) \leqslant-\gamma \leqslant  \min (\eta U_{21}^d+\beta U_{21}^a,\eta U_{22}^d+\beta U_{22}^a). 
$ 
Similarly, we can get $
\max (\eta U_{21}^d+\beta U_{21}^a,\eta U_{22}^d+\beta U_{22}^a) \leqslant-\gamma \leqslant  \min (\eta U_{11}^d +\beta U_{11}^a, \eta U_{12}^d+\beta U_{12}^a), 
$ if $\phi<0$.

Necessity: Consider 
$   \max (\eta U_{11}^d +\beta U_{11}^a, \eta U_{12}^d+\beta U_{12}^a) \leqslant -\gamma\leqslant \min (\eta U_{21}^d+\beta U_{21}^d,\eta U_{22}^d+\beta U_{22}^a)$. Then we have (\ref{eq::zhuanhua}). If all inequalities in (\ref{eq::zhuanhua}) are not strctly satisfied, then the ZD strategy enforces  $\eta U_d+\beta U_a+\gamma=0$. Otherwise, take $\phi = \max\{|\eta U_{i}^d+\beta U_i^a+\gamma|\}_{i\in\{11,12,21,22\}}$, and we obtain
$$
\begin{aligned}
-1\leqslant \frac{1}{\phi}(\eta U_{i}^d + \beta U_{i}^a + \gamma ) \leqslant 0,i\in\{11,12\},\\
0\leqslant \frac{1}{\phi}(\eta U_{j}^d + \beta U_{j}^a + \gamma ) \leqslant 1, j\in\{21,22\}.
\end{aligned}
$$
Therefore, the ZD strategy $\pi_d^{ZD}(\frac{\eta}{\phi},\frac{\beta}{\phi},\frac{\gamma}{\phi})$ is feasible, and it enforces  $\eta U_d+\beta U_a+\gamma=0$. Similarly, the conclusion holds for $\max\limits_{s\in\{21,22\}} \eta U_{s}^d+\beta U_{s}^a \leqslant-\gamma \leqslant   \min\limits_{s\in\{11,12\}} \eta U_{s}^d +\beta U_{s}^a.$

\section{\label{app::T1}Proof of Theorem 1}

Consider $(U_{21}^d,U_{21}^a) ,(U_{22}^d,U_{22}^a) \in \Gamma^{+}(U_{11}^d,U_{11}^a,U_{12}^d,U_{12}^a)$. Take $\eta=-\frac{U_{21}^a-U_{11}^a}{U_{21}^d-U_{11}^d}$, $\beta=1$, $\gamma=U_{11}^d\frac{U_{21}^a-U_{11}^a}{U_{21}^d-U_{11}^d}-U_{11}^a$, and we have $$
\max (\eta U_{11}^a +\beta U_{11}^b, \eta U_{12}^a+\beta U_{12}^b) \leqslant-\gamma \leqslant  \min (\eta U_{21}^a+\beta U_{21}^b,\eta U_{22}^a+\beta U_{22}^b). 
$$
Similarly, the conclusion holds for $(U_{21}^d,U_{21}^a) ,(U_{22}^d,U_{22}^a) \in \Gamma^{-}(U_{11}^d,U_{11}^a,U_{12}^d,U_{12}^a)$.  Thus, there exists at least a ZD strategy for the defender.

\section{\label{app::T2}Proof of Theorem 2}

1) When $U_{11}^d\geqslant U_{21}^d$ and $U_{11}^a\geqslant U_{21}^a$, the ZD strategy $\pi^{ZD}(-k_1, 1,k_1U_{11}^d - U_{11}^a)$ is feasible for the defender according to Lemma 1, where  $0\leqslant k_1\leqslant \frac{U_{11}^a-U_{21}^a}{U_{11}^d-U_{21}^d} $. The ZD strategy enforces  $U_a-U_{11}^a=k_1(U_d-U_{11}^d)$. Since $k_1\geqslant0$, the  optimal utility of the attacker is  $U_{11}^a$ when the attacker observes the defender's strategy. In this case, the defender's utility is $U_{11}^d$. Notice that $U_{11}^d$ is also the optimal for the defender since $U_{11}^d\leqslant\max\{U_{12}^d,U_{21}^d,U_{22}^d\}$. Then  $U_d^{SSE}\leqslant U_{11}^d$. According to Lemma 2, $U_d^{SSE}=U_d(\pi_d^{ZD}, \pi_a^{BR}(\pi_d^{ZD}))$. 

2) When $U_{11}^d<U_{21}^d$ and $U_{11}^a\geqslant U_{21}^a$, the ZD strategy $\pi^{ZD}(0,1,-U_{21}^a)$ is also feasible for the defender according to Lemma 1. The ZD strategy enforces $U_a-U_{21}^a=0$. Since the attacker always breaks ties optimally for the defender if there are multiple options, the attacker chooses the strategy which enforces $U_{a}=U_{21}^a $ and $U_{d}=\frac{(U_{21}^a-U_{11}^a)(U_{22}^d-U_{11}^d)}{U_{22}^2-U_{11}^2}+U_{11}^d$. Thus, $U_d^{SSE}\geqslant \frac{(U_{21}^a-U_{11}^a)(U_{22}^d-U_{11}^d)}{U_{22}^2-U_{11}^2}+U_{11}^d$.  Further, suppose  that the defender's utility is higher than $\frac{(U_{21}^a-U_{11}^a)(U_{22}^d-U_{11}^d)}{U_{22}^2-U_{11}^2}+U_{11}^d$, when it chooses SSE strategy, i.e., $U_d^{SSE}>\frac{(U_{21}^a-U_{11}^a)(U_{22}^d-U_{11}^d)}{U_{22}^2-U_{11}^2}+U_{11}^d$. Since $U_a^{SSE}-U_{11}^a\leqslant \frac{(U_{22}^a-U_{11}^a)(U_d^{SSE}-U_{11}^d)}{U_{22}^d-U_{11}^d}$ and $\frac{U_{22}^a-U_{11}^a}{U_{22}^d-U_{11}^d}<0$, we have $U_a^{SSE}<U_{12}^a$. Actually, for any defender's strategy $\pi_d$, when the attacker chooses the strategy $\pi_a$ with $\pi_a(1|11)=\pi_a(1|21)=\pi_a(2|12)=\pi_a(2|22)=1$, $U_a(\pi_d,\pi_a)=U_{11}^a\frac{\pi_d(1|21)}{\pi_d(2|11)+\pi_d(1|21)}+U_{21}^a\frac{\pi_d(2|11)}{\pi_d(2|11)+\pi_d(1|21)}\geqslant U_{21}^a$. The attacker always has a strategy to get a utility no lower than $U_{12}^a$, which is conflict with $U_{a}^{SSE}<U_{12}^a$. Thus, $U_d^{SSE}= \frac{(U_{21}^a-U_{11}^a)(U_{22}^d-U_{11}^d)}{U_{22}^2-U_{11}^2}+U_{11}^d=U_d(\pi_d^{ZD}, \pi_a^{BR}(\pi_d^{ZD})$.

3) When $ U_{11}^a<U_{21}^a$,  the ZD strategy $\pi^{ZD}(-k_2,1, k_2 U_{12}^d-U_{12}^a)$ is feasible for the defender according to Lemma 1, where  $\frac{U_{22}^a-U_{12}^a}{U_{22}^d-U_{12}^d} \leqslant k_2\leqslant\frac{U_{11}^a-U_{12}^a}{U_{11}^d-U_{12}^d}$. Thus, after observing the defender's strategy, the optimal utility for the attacker is $U_{12}^a$. In this case, the defender's corresponding utility is $U_{12}^d$, and  $U_d(\pi_d^{ZD}, \pi_a^{BR}(\pi_d^{ZD}))-U_d(\pi_d^{SSE}, \pi_a^{BR}(\pi_d^{SSE})= U_d^{SSE}-U_{12}^d$. Moreover, according to Lemma 1, for any  ZD strategy $\pi_d$ that enforces $\eta U_d+ \beta U_a+\gamma=0$, $\eta\cdot \beta>0$ always holds when $ U_{11}^a<U_{21}^a$. Thus, the attacker's BR strategy also minimizes the defender's utility, and $\max\limits_{\pi_d^{ZD}\in \Xi }U_d(\pi_d^{ZD}, \pi_a^{BR}(\pi_d^{ZD}))=U_{12}^d$. Then $\min\limits_{\pi_d^{ZD}\in \Xi }U_d(\pi_d^{SSE}, \pi_a^{BR}(\pi_d^{SSE}))-U_d(\pi_d^{ZD}, \pi_a^{BR}(\pi_d^{ZD}))=U_d^{SSE}-U_{12}^d.$

Therefore, $\begin{aligned}&\min\limits_{\pi_d^{ZD}\in \Xi}U_d\left(\pi_d^{SSE}, \pi_a^{BR}(\pi_d^{SSE})\right)\!-\!U_d\left(\pi_d^{ZD}, \pi_a^{BR}(\pi_d^{ZD})\right)
= \left\{
\begin{array}{ll}
0, &\quad \text{if }  U_{11}^d\geqslant U_{21}^d,\\
U_d^{SSE}-U_{12}^d, &\quad \text{if }U_{11}^d < U_{21}^d.
\end{array}
\right.
\end{aligned}$

\section{\label{app::T3}Proof of Theorem 3}
\iffalse
 For any $\mathbf{f}=[f_1,f_2,f_3,f_4]^T\in\mathbb{R}^4$, take
$$
\begin{aligned}
D(\pi_d,\pi_a,\mathbf{f})=det\left[\begin{array}{llll}
\pi_d(1|11)\pi_a(1|11)-1 & \pi_d(1|11)-1 & \pi_a(1|11) -1&f_1\\
\pi_d(1|12)\pi_a(1|12) &\pi_d(1|12)-1  &\pi_a(1|12)  &f_2 \\
\pi_d(1|21)\pi_a(1|21) &\pi_d(1|21) &\pi_a(1|21)-1  &f_3 \\
\pi_d(1|22)\pi_a(1|22) &\pi_d(1|22)  &\pi_a(1|22)  &f_4 
\end{array}\right].
\end{aligned}
$$\fi
According to \cite{press2012iterated}, $U_d(\pi_d,\pi_a)= \frac{D(\pi_d,\pi_a,S^d)}{D(\pi_d,\pi_a,\mathbf{1})}$ and $U_a(\pi_d,\pi_a)= \frac{D(\pi_d,\pi_a,S^a)}{D(\pi_d,\pi_a,\mathbf{1})}$. For the stubborn attacker with $\pi_a^*(1|11)=\pi_a^*(1|21)=1$,  we have
$$U_d(\pi_d,\pi_a^*)=U_{11}^d\frac{\pi_d(1|21)}{\pi_d(2|11)+\pi_d(1|21)}+U_{21}^d\frac{\pi_d(2|11)}{\pi_d(2|11)+\pi_d(1|21)},$$ and 
$$ U_a(\pi_d,\pi_a^*)=U_{11}^a\frac{\pi_d(1|21)}{\pi_d(2|11)+\pi_d(1|21)}+U_{21}^a\frac{\pi_d(2|11)}{\pi_d(2|11)+\pi_d(1|21)},$$ for any $\pi_a^*(1|s)=1$ or $0$, where $s\in\{12,22\}$. Notice that $U_d(\pi_d,\pi_a^*)$ and $U_a(\pi_d,\pi_a^*)$ are monotonous in $\pi_a^*(1|s)\in[0,1]$. Then  $U_d(\pi_d^{SSE},\pi_a^*)=\frac{U_{11}^d \pi_d^{SSE}(1|21)+U_{21}^d\pi_d^{SSE}(2|11)}{\pi_d^{SSE}(2|11)+\pi_d^{SSE}(1|21)}$. Take $\pi_d^{ZD}= \pi^{ZD}(-k, 1, kU_{12}^d -U_{12}^a)$ with $k=\frac{U_{11}^a-U_{12}^a}{U_{11}^d-U_{12}^d}$, and we have $U_d(\pi_d^{ZD},\pi_a^*)=U_{11}^d$, which implies
$U_d(\pi_d^{ZD},\pi_a^*)- U_d(\pi_d^{SSE},\pi_a^*)=U_{11}^d-\frac{U_{11}^d \pi_d^{SSE}(1|21)+U_{21}^d\pi_d^{SSE}(2|11)}{\pi_d^{SSE}(2|11)+\pi_d^{SSE}(1|21)}\geqslant 0.$

\iffalse
$$ \begin{aligned}
D(\pi_d,\pi_a,\mathbf{f})=&det\left[\begin{array}{llll}
\pi_d(1|11)-1 & \pi_d(1|11)-1 & 0&f_1\\
\pi_d(1|12)\pi_a(1|12) &\pi_d(1|12)-1  &\pi_a(1|12)  &f_2 \\
\pi_d(1|21) &\pi_d(1|21) &0 &f_3 \\
\pi_d(1|22)\pi_a(1|22) &\pi_d(1|22)  &\pi_a(1|22)  &f_4 
\end{array}\right].\\
=& f_1
\end{aligned}
$$
\fi

\section{\label{app::T4}Proof of Theorem 4}
If $U_{11}^d\!\geqslant\! U_{22}^d$ and $ U_{11}^a\!\geqslant \!U_{21}^a$, the ZD strategy $\pi^{ZD}(-k_1,\! 1,\! k_1U_{11}^d \!-\!U_{11}^a)$ is feasible for the defender according to Lemma 1. Moreover, according to Theorem 2, this ZD strategy is also an SSE strategy. Therefore, this ZD strategy brings the defender the same utility as the SSE strategy against a boundedly rational attacker. Thus, 
$$\min\limits_{\pi_d^{ZD}\in \Xi}\! U_d\left(\pi_d^{SSE}\!,\!\pi_a^{\lambda}(\pi_d^{SSE}\!,\!\pi_a^*)\!\right)\!- \!U_d\left(\pi_d^{ZD}\!,\!\pi_a^{\lambda}(\pi_d^{ZD},\!\pi_a^*)\!\right) \leqslant 0, \ \text{ if } \ U_{11}^d\!\geqslant\! U_{22}^d, \ U_{11}^a\!\geqslant \!U_{21}^a.
$$ 

Otherwise, the ZD strategy $\pi^{ZD}(-k_2,\!1, \!k_2 U_{12}^d\!-\!U_{12}^a)$ is feasible for the defender by Lemma 1. According to Theorem 2, $U_d(\pi_d^{ZD}, \pi_a^{BR}(\pi_d^{ZD}))-U_d(\pi_d^{SSE}, \pi_a^{BR}(\pi_d^{SSE}))= -(U_d^{SSE}-U_{12}^d).$ Then
$$\begin{aligned}
\frac{D(\pi_d^{ZD}, \pi_a^{BR}(\pi_d^{ZD}),S^d)}{D(\pi_d^{ZD}, \pi_a^{BR}(\pi_d^{ZD}),\textbf{1})}-\frac{D(\pi_d^{SSE}, \pi_a^{BR}(\pi_d^{SSE}),S^d)}{D(\pi_d^{SSE}, \pi_a^{BR}(\pi_d^{SSE}),\textbf{1})}=-(U_d^{SSE}-U_{12}^d).
\end{aligned}
$$ As a result, 
\begin{equation}\label{eq::SSEZD-1}\begin{aligned}&D(\pi_d^{ZD}, \pi_a^{BR}(\pi_d^{ZD}),S^d)D(\pi_d^{SSE}, \pi_a^{BR}(\pi_d^{SSE}),\textbf{1})-D(\pi_d^{SSE}, \pi_a^{BR}(\pi_d^{SSE}),S^d)D(\pi_d^{ZD}, \pi_a^{BR}(\pi_d^{ZD}),\textbf{1})\\
= & -(U_d^{SSE}-U_{12}^d)D(\pi_d^{ZD}, \pi_a^{BR}(\pi_d^{ZD}),\textbf{1})D(\pi_d^{SSE}, \pi_a^{BR}(\pi_d^{SSE}),\textbf{1})\end{aligned}\end{equation}
 
It follows from Theorem 3 that $U_d(\pi_d^{ZD},\pi_a^*)-U_d(\pi_d^{SSE},\pi_a^*)= U_{11}^a-\frac{U_{11}^a \pi_d^{SSE}(1|21)+U_{21}^a\pi_d^{SSE}(2|11)}{\pi_d^{SSE}(2|11)+\pi_d^{SSE}(1|21)}$. Similarly, 
\begin{equation}\label{eq::SSEZD-2}\frac{D(\pi_d^{ZD}, \pi_a^*,S^d)}{D(\pi_d^{ZD},\pi_a^*,\textbf{1})}-\frac{D(\pi_d^{SSE}, \pi_a^*,S^d)}{D(\pi_d^{SSE}, \pi_a^*,\textbf{1})}= U_{11}^a-\frac{U_{11}^a \pi_d^{SSE}(1|21)+U_{21}^a\pi_d^{SSE}(2|11)}{\pi_d^{SSE}(2|11)+\pi_d^{SSE}(1|21)}.\end{equation}
Recall  
$$
U_d(\pi_d^{ZD},\lambda\pi_a^{BR}(\pi_d^{ZD})+(1-\lambda)\pi_a^*)=\frac{D(\pi_d^{ZD},\lambda\pi_a^{BR}(\pi_d^{ZD})+(1-\lambda)\pi_a^*,S^d)}{D(\pi_d^{ZD},\lambda\pi_a^{BR}(\pi_d^{ZD})+(1-\lambda)\pi_a^*,\textbf{1})},
$$
$$
 U_d(\pi_d^{SSE},\lambda\pi_a^{BR}(\pi_d^{SSE})+(1-\lambda)\pi_a^*)=\frac{D(\pi_d^{SSE},\lambda\pi_a^{BR}(\pi_d^{SSE})+(1-\lambda)\pi_a^*,S^d)}{D(\pi_d^{SSE},\lambda\pi_a^{BR}(\pi_d^{SSE})+(1-\lambda)\pi_a^*,\textbf{1})}. 
$$

%Take $C(\pi_d^{ZD},\pi_d^{SSE},\pi_a^*,\lambda)=D(\pi_d^{ZD},\lambda\pi_a^{BR}(\pi_d^{ZD})+(1-\lambda)\pi_a^*,\textbf{1})D(\pi_d^{SSE},\lambda\pi_a^{BR}(\pi_d^{SSE})+(1-\lambda)\pi_a^*,\textbf{1}) $.

 According to \cite{press2012iterated}, $C(\pi_d^{ZD},\pi_d^{SSE},\pi_a^*,\lambda)\neq 0$, which was defined in Appendix \ref{app::No}. Without loss of generality, we consider $C(\pi_d^{ZD},\pi_d^{SSE},\pi_a^*,\lambda)>0$.   As a result,
$$
\begin{aligned}
&U_d(\pi_d^{SSE},\lambda\pi_a^{BR}(\pi_d^{SSE})+(1-\lambda)\pi_a^*)-U_d(\pi_d^{ZD},\lambda\pi_a^{BR}(\pi_d^{ZD})+(1-\lambda)\pi_a^*)\\
=& \frac{D(\pi_d^{SSE},\lambda\pi_a^{BR}(\pi_d^{SSE})+(1-\lambda)\pi_a^*,S^d)}{D(\pi_d^{SSE},\lambda\pi_a^{BR}(\pi_d^{SSE})+(1-\lambda)\pi_a^*,\textbf{1})}-\frac{D(\pi_d^{ZD},\lambda\pi_a^{BR}(\pi_d^{ZD})+(1-\lambda)\pi_a^*,S^d)}{D(\pi_d^{ZD},\lambda\pi_a^{BR}(\pi_d^{ZD})+(1-\lambda)\pi_a^*,\textbf{1})}\\
=&\frac{1}{C(\pi_d^{ZD},\pi_d^{SSE},\pi_a^*,\lambda)} \left( D(\pi_d^{SSE},\lambda\pi_a^{BR}(\pi_d^{SSE})+(1-\lambda)\pi_a^*,S^d)D(\pi_d^{ZD},\lambda\pi_a^{BR}(\pi_d^{ZD})+(1-\lambda)\pi_a^*,\textbf{1})\right.\\
&\left.-D(\pi_d^{ZD},\lambda\pi_a^{BR}(\pi_d^{ZD})+(1-\lambda)\pi_a^*,S^d)D(\pi_d^{SSE},\lambda\pi_a^{BR}(\pi_d^{SSE})+(1-\lambda)\pi_a^*,\textbf{1})\right).
\end{aligned}
$$

Actually, for any $\pi_d\in\Delta\mathcal{D},\pi_a^1,\pi_a^2\in\Delta\mathcal{A}, \lambda\in[0,1]$, and $\mathbf{f}=[f_1,f_2,f_3,f_4]^T\in\mathbb{R}^4$,
\iffalse
the following equation holds.
$$D(\pi_a,\lambda \pi_b^1+(1-\lambda)\pi_b^2,f)=\lambda^2 D(\pi_a,\pi_b^1,f) + (1-\lambda)^2  D(\pi_a,\pi_b^2,f)+(\lambda)(1-\lambda)(D(\pi_a,\pi_b^1,\pi_b^2,f)+D(\pi_a,\pi_b^2,\pi_b^1,f)).$$
That is bacuse 
\fi

$$\begin{aligned}&{D(\pi_d,\lambda\pi_a^1+(1-\lambda)\pi_a^2,\mathbf{f})}\\
=&det\left[\begin{array}{llll}
\pi_d(1|11)(\lambda\pi_a^1(1|11)+(1-\lambda)\pi_a^2(1|11))-1  & \pi_d(1|11)-1 &\lambda\pi_a^1(1|11)+(1-\lambda)\pi_a^2(1|11) -1&f_1\\
\pi_d(1|12)(\lambda\pi_a^1(1|12)+(1-\lambda)\pi_a^2(1|12)) & \pi_d(1|12)-1 & \lambda\pi_a^1(1|12)+(1-\lambda)\pi_a^2(1|12)&f_2 \\
\pi_d(1|21)(\lambda\pi_a^1(1|21)+(1-\lambda)\pi_a^2(1|21)) & \pi_d(1|21)&  \lambda\pi_a^1(1|21)+(1-\lambda)\pi_a^2(1|21) -1&f_3\\
\pi_d(1|22)(\lambda\pi_a^1(1|22)+(1-\lambda)\pi_a^2(1|22)) & \pi_d(1|22) &\lambda\pi_a^1(1|22)+(1-\lambda)\pi_a^2(1|22) & f_4
\end{array}\right]\\
=&det\left[\begin{array}{llll}
\pi_d(1|11)(\lambda\pi_a^1(1|11)+(1-\lambda)\pi_a^2(1|11))-1  & \pi_d(1|11)-1 &\lambda\pi_a^1(1|11) -\lambda &f_1\\
\pi_d(1|12)(\lambda\pi_a^1(1|12)+(1-\lambda)\pi_a^2(1|12)) & \pi_d(1|12)-1 & \lambda\pi_a^1(1|12)&f_2 \\
\pi_d(1|21)(\lambda\pi_a^1(1|21)+(1-\lambda)\pi_a^2(1|21)) & \pi_d(1|21)&  \lambda\pi_a^1(1|21)-\lambda&f_3\\
\pi_d(1|22)(\lambda\pi_a^1(1|22)+(1-\lambda)\pi_a^2(1|22)) & \pi_d(1|22) &\lambda\pi_a^1(1|22) & f_4
\end{array}\right]\\
&+det\left[\begin{array}{llll}
\pi_d(1|11)(\lambda\pi_a^1(1|11)+(1-\lambda)\pi_a^2(1|11))-1  & \pi_d(1|11)-1 &(1-\lambda)\pi_a^2(1|11) -(1-\lambda)&f_1\\
\pi_d(1|12)(\lambda\pi_a^1(1|12)+(1-\lambda)\pi_a^2(1|12)) & \pi_d(1|12)-1 & (1-\lambda)\pi_a^2(1|12)&f_2 \\
\pi_d(1|21)(\lambda\pi_a^1(1|21)+(1-\lambda)\pi_a^2(1|21)) & \pi_d(1|21)& (1-\lambda)\pi_a^2(1|21) -(1-\lambda)&f_3\\
\pi_d(1|22)(\lambda\pi_a^1(1|22)+(1-\lambda)\pi_a^2(1|22)) & \pi_d(1|22) &(1-\lambda)\pi_a^2(1|22) & f_4
\end{array}\right]\\
=&\lambda^2 D(\pi_d,\pi_a^1,\mathbf{f}) + (1-\lambda)^2  D(\pi_d,\pi_a^2,\mathbf{f})+\lambda(1-\lambda)(D(\pi_d,\pi_a^1,\pi_a^2,\mathbf{f})+D(\pi_d,\pi_a^2,\pi_a^1,\mathbf{f})),
\end{aligned}$$
where $D(\pi_d,\pi_a^1,\pi_a^2,\mathbf{f})$ is shown in (\ref{eq::D4}). 
\iffalse $$D(\pi_d,\pi_a^1,\pi_a^2,f)=\left[\begin{array}{llll}
\pi_d(1|11)\pi_a^2(1|11)-1  & \pi_d(1|11)-1 &\pi_a^1(1|11) -1 &f_1\\
\pi_d(1|12)\pi_a^2(1|12) & \pi_d(1|12)-1 & \pi_a^1(1|12)&f_2 \\
\pi_d(1|21)\pi_a^2(1|21) & \pi_d(1|21)&  \pi_a^1(1|21)-1&f_3\\
\pi_d(1|22)\pi_a^2(1|22) & \pi_d(1|22) &\pi_a^1(1|22) & f_4
\end{array}\right].$$\fi

Based on the above equation, we obtain 
\begin{equation}
\begin{aligned}
\!\!\!\!& D(\pi_d^{ZD},\lambda\pi_a^{BR}(\pi_d^{ZD})+(1-\lambda)\pi_a^*,S^d)D(\pi_d^{SSE},\lambda\pi_a^{BR}(\pi_d^{SSE})+(1-\lambda)\pi_a^*,\textbf{1})\\
= &\left( \lambda^2 D(\pi_d^{ZD},\pi_a^{BR}(\pi_d^{ZD}),S^d) + (1-\lambda)^2 D(\pi_d^{ZD},\pi_a^*,S^d)+\lambda(1-\lambda)(D(\pi_d^{ZD},\pi_a^{BR}(\pi_d^{ZD}),\pi_a^*,S^d)\right.\\
& \left. + D(\pi_d^{ZD},\pi_a^*,\pi_a^{BR}(\pi_d^{ZD}),S^d) \right)\times\left( \lambda^2 D(\pi_d^{SSE},\pi_a^{BR}(\pi_d^{SSE}),\textbf{1})  + (1-\lambda)^2 D(\pi_d^{SSE},\pi_a^*,\textbf{1})\right. \\
&\left.+ \lambda(1-\lambda)(D(\pi_d^{SSE},\pi_a^{BR}(\pi_d^{SSE}),\pi_a^*,\textbf{1})+D(\pi_d^{SSE},\pi_a^*,\pi_a^{BR}(\pi_d^{SSE}),\textbf{1})  \right)\\
=&\lambda^4 D(\pi_d^{ZD},\pi_a^{BR}(\pi_d^{ZD}),S^d)D(\pi_d^{SSE},\pi_a^{BR}(\pi_d^{SSE}),\textbf{1})+(1-\lambda)^4 D(\pi_d^{ZD},\pi_a^*,S^d)D(\pi_d^{SSE},\pi_a^*,\textbf{1}) \\
& +\!\lambda^2(1\!-\!\lambda)^2\! D(\pi_d^{ZD}\!\!,\pi_a^{BR}\!(\pi_d^{ZD}\!),S^d) \!D(\pi_d^{SSE}\!,\pi_a^*\!,\textbf{1})\!+\!\lambda^2(1\!-\!\lambda)^2 \! D(\pi_d^{SSE}\!,\pi_a^{BR}(\pi_d^{SSE}\!),S^d)\! D(\pi_d^{ZD}\!,\pi_a^*\!,\textbf{1})\\
&+\lambda^3(1-\lambda)\left( D(\pi_d^{ZD},\pi_a^{BR}(\pi_d^{ZD}),S^d) J(\pi_d^{SSE},\pi_a^*,\textbf{1}) + D(\pi_d^{SSE},\pi_a^{BR}(\pi_d^{SSE}),\textbf{1})J(\pi_d^{ZD},\pi_a^*,S^d)\right) \\
&+\lambda(1-\lambda)^3\left( D(\pi_d^{ZD},\pi_a^*,S^d) J(\pi_d^{SSE},\pi_a^*,\textbf{1}) + D(\pi_d^{SSE},\pi_a^*,\textbf{1})J(\pi_d^{ZD},\pi_a^*,S^d)\right) \\
&+\lambda^2(1-\lambda)^2J(\pi_d^{ZD},\pi_a^*,S^d)J(\pi_d^{SSE},\pi_a^*,\textbf{1}).
\end{aligned}
\end{equation}
Similarly, 
\begin{equation}
\begin{aligned}
& D(\pi_d^{SSE},\lambda\pi_a^{BR}(\pi_d^{SSE})+(1-\lambda)\pi_a^*,S^d)D(\pi_d^{ZD},\lambda\pi_a^{BR}(\pi_d^{ZD})+(1-\lambda)\pi_a^*,\textbf{1})\\
= &\left( \lambda^2 D(\pi_d^{SSE},\pi_a^{BR}(\pi_d^{SSE}),S^d) + (1-\lambda)^2 D(\pi_d^{SSE},\pi_a^*,S^d)+\lambda(1-\lambda)(D(\pi_d^{SSE},\pi_a^{BR}(\pi_d^{SSE}),\pi_a^*,S^d)\right.\\
&\left. +D(\pi_d^{SSE},\pi_a^*,\pi_a^{BR}(\pi_d^{SSE}),S^d) \right)\times\left( \lambda^2 D(\pi_d^{ZD},\pi_a^{BR}(\pi_a^{ZD}),\textbf{1}) + (1-\lambda)^2 D(\pi_d^{ZD},\pi_a^*,\textbf{1})\right. \\
&\left.+ \lambda(1-\lambda)(D(\pi_d^{ZD},\pi_a^{BR}(\pi_d^{ZD}),\pi_a^*,\textbf{1})+D(\pi_d^{ZD},\pi_a^*,\pi_a^{BR}(\pi_d^{ZD}),\textbf{1})  \right)\\
=&\lambda^4 D(\pi_d^{SSE},\pi_a^{BR}(\pi_d^{SSE}),S^d)D(\pi_d^{ZD},\pi_a^{BR}(\pi_d^{ZD}),\textbf{1})+(1-\lambda)^4 D(\pi_d^{SSE},\pi_a^*,S^d)D(\pi_d^{ZD},\pi_a^*,\textbf{1}) \\
& +\lambda^2(1-\lambda)^2 D(\pi_d^{SSE},\pi_a^{BR}(\pi_d^{SSE}),S^d) D(\pi_d^{ZD},\pi_a^*,\textbf{1})+\lambda^2(1-\lambda)^2  D(\pi_d^{ZD},\pi_a^{BR}(\pi_d^{ZD}),S^d) D(\pi_d^{SSE},\pi_a^*,\textbf{1})\\
&+\lambda^3(1-\lambda)\left( D(\pi_d^{SSE},\pi_a^{BR}(\pi_d^{SSE}),S^d) J(\pi_d^{ZD},\pi_a^*,\textbf{1}) + D(\pi_d^{ZD},\pi_a^{BR}(\pi_d^{ZD}),\textbf{1})J(\pi_d^{SSE},\pi_a^*,S^d)\right) \\
&+\lambda(1-\lambda)^3\left( D(\pi_d^{SSE},\pi_a^*,S^d) J(\pi_d^{ZD},\pi_a^*,\textbf{1}) + D(\pi_d^{ZD},\pi_a^*,\textbf{1})J(\pi_d^{SSE},\pi_a^*,S^d)\right) \\
&+\lambda^2(1-\lambda)^2J(\pi_d^{SSE},\pi_a^*,S^d)J(\pi_d^{ZD},\pi_a^*,\textbf{1}).
\end{aligned}
\end{equation}
By taking the subtraction  between the above two equations,
$$
\begin{aligned}
& D(\pi_d^{SSE},\lambda\pi_a^{BR}(\pi_d^{SSE})+(1-\lambda)\pi_a^*,S^d)D(\pi_d^{ZD},\lambda\pi_a^{BR}(\pi_d^{ZD})+(1-\lambda)\pi_a^*,\textbf{1})\\
&- D(\pi_d^{ZD},\lambda\pi_a^{BR}(\pi_d^{ZD})+(1-\lambda)\pi_a^*,S^d)D(\pi_d^{SSE},\lambda\pi_a^{BR}(\pi_d^{SSE})+(1-\lambda)\pi_a^*,\textbf{1})\\
=& \lambda^4\left( D(\pi_d^{SSE},\pi_a^{BR}(\pi_d^{SSE}),S^d)D(\pi_d^{ZD},\pi_a^{BR}(\pi_d^{ZD}),\textbf{1})-D(\pi_d^{ZD},\pi_a^{BR}(\pi_d^{ZD}),S^d)D(\pi_d^{SSE},\pi_a^{BR}(\pi_d^{SSE}),\textbf{1}) \right)\\
& + (1-\lambda)^4 \left(D(\pi_d^{SSE},\pi_a^*,S^d)D(\pi_d^{ZD},\pi_a^*,\textbf{1})- D(\pi_d^{ZD},\pi_a^*,S^d)D(\pi_d^{SSE},\pi_a^*,\textbf{1})\right)-g,
\end{aligned}
$$
where 
\begin{equation}\label{eq::gdef}
\begin{aligned}
g=&\lambda^3(1-\lambda)\left( D(\pi_d^{ZD},\pi_a^{BR}(\pi_d^{ZD}),S^d) J(\pi_d^{SSE},\pi_a^*,\textbf{1}) + D(\pi_d^{SSE},\pi_a^{BR}(\pi_d^{SSE}),\textbf{1})J(\pi_d^{ZD},\pi_a^*,S^d)\right) \\
&-\lambda^3(1-\lambda)\left( D(\pi_d^{SSE},\pi_a^{BR}(\pi_d^{SSE}),S^d) J(\pi_d^{ZD},\pi_a^*,\textbf{1}) + D(\pi_d^{ZD},\pi_a^{BR}(\pi_d^{ZD}),\textbf{1})J(\pi_d^{SSE},\pi_a^*,S^d)\right) \\
&+\lambda(1-\lambda)^3\left( D(\pi_d^{ZD},\pi_a^*,S^d) J(\pi_d^{SSE},\pi_a^*,\textbf{1}) + D(\pi_d^{SSE},\pi_a^*,\textbf{1})J(\pi_d^{ZD},\pi_a^*,S^d)\right) \\
&-\lambda(1-\lambda)^3\left( D(\pi_d^{SSE},\pi_a^*,S^d) J(\pi_d^{ZD},\pi_a^*,\textbf{1}) + D(\pi_d^{ZD},\pi_a^*,\textbf{1})J(\pi_d^{SSE},\pi_a^*,S^d)\right) \\
&+\lambda^2(1-\lambda)^2J(\pi_d^{ZD},\pi_a^*,S^d)J(\pi_d^{SSE},\pi_a^*,\textbf{1})\\
&-\lambda^2(1-\lambda)^2J(\pi_d^{SSE},\pi_a^*,S^d)J(\pi_d^{ZD},\pi_a^*,\textbf{1}).
\end{aligned}
\end{equation}
%Notice the defination of $B_1,B_2,B_3$, and $B$ in 

Recall the definitions of $B_1,B_2,B_3$, and $B$ in Appendix \ref{app::No},
\iffalse
For the sake of  analysis, we take
$$
\begin{aligned}
B_1=&\max\limits_{\pi_d^{SSE},\pi_d^{ZD}}\max\limits_{f\in\{\textbf{1},S^d\}}\frac{1}{2}\left|D(\pi_d^{ZD},\pi_a^{BR}(\pi_d^{ZD}),S^d) D(\pi_d^{SSE},f)-  D(\pi_d^{SSE},\pi_a^{BR}(\pi_d^{SSE}),S^d) D(\pi_d^{ZD},f)\right|,\\
B_2=&\max\limits_{\pi_d^{SSE},\pi_d^{ZD}}\left| D(\pi_d^{ZD},\pi_a^*,S^d) J(\pi_d^{SSE},\pi_a^*,\textbf{1}) + D(\pi_d^{SSE},\pi_a^*,\textbf{1})J(\pi_d^{ZD},\pi_a^*,S^d)- D(\pi_d^{SSE},\pi_a^*,S^d) J(\pi_d^{ZD},\pi_a^*,\textbf{1})\right. \\
& \left. \quad \quad\quad\quad  - D(\pi_d^{ZD},\pi_a^*,\textbf{1})J(\pi_d^{SSE},\pi_a^*,S^d)\right|,\\
B_3=&\max_{\pi_d^{SSE},\pi_d^{ZD}}\left|J(\pi_d^{ZD},\pi_a^*,S^d)J(\pi_d^{SSE},\pi_a^*,\textbf{1})- J(\pi_d^{SSE},\pi_a^*,S^d)J(\pi_d^{ZD},\pi_a^*,\textbf{1})\right|,\\
B=&\max\left\{B_1,B_2,\frac{1}{2}B_3\right\}.
\end{aligned}
$$\fi
and  we have
\begin{equation}\label{eq::g}
\begin{aligned}
|g|\leqslant& B\lambda^3(1-\lambda)+B\lambda(1-\lambda)^3+2B\lambda(1-\lambda)^2\\
=&B\lambda(1-\lambda)(\lambda^2+(1-\lambda)^2+2\lambda(1-\lambda))\\
=&B\lambda(1-\lambda)(\lambda+(1-\lambda))^2\\
=&B\lambda(1-\lambda).
\end{aligned}
\end{equation}
Recall (\ref{eq::SSEZD-1}) and (\ref{eq::SSEZD-2}), and we obtain 
$$
\begin{aligned}
& D(\pi_d^{SSE},\lambda\pi_a^{BR}(\pi_d^{SSE})+(1-\lambda)\pi_a^*,S^d)D(\pi_d^{ZD},\lambda\pi_a^{BR}(\pi_d^{ZD})+(1-\lambda)\pi_a^*,\textbf{1})\\
&- D(\pi_d^{ZD},\lambda\pi_a^{BR}(\pi_d^{ZD})+(1-\lambda)\pi_a^*,S^d)D(\pi_d^{SSE},\lambda\pi_a^{BR}(\pi_d^{SSE})+(1-\lambda)\pi_a^*,\textbf{1})\\
\geqslant & \lambda^4 \left(U_d^{SSE}-U_{12}^d\right)D(\pi_d^{ZD}, \pi_a^{BR}(\pi_d^{ZD}),\textbf{1})D(\pi_d^{SSE}, \pi_a^{BR}(\pi_d^{SSE}),\textbf{1} )\\
& -(1-\lambda)^4 \left(U_{11}^d-\frac{U_{11}^d \pi_d^{SSE}(1|21)+U_{21}^d\pi_d^{SSE}(2|11)}{\pi_d^{SSE}(2|11)+\pi_d^{SSE}(1|21)} \right)D(\pi_d^{ZD}, \pi_a^*,\textbf{1})D(\pi_d^{SSE}, \pi_a^*,\textbf{1}) \\
&\left.-B\lambda(1-\lambda)\right).
\end{aligned}
$$
Actually, $D(\pi_d^{ZD}, \pi_a^*,\textbf{1})=\frac{1\cdot \pi_d^{ZD}(1|21)+1\cdot \pi_d^{ZD}(2|11)}{\pi_d^{ZD}(2|11)+\pi_d^{ZD}(1|21)}=1,$ and $D(\pi_d^{SSE}, \pi_a^*,\textbf{1})= \frac{1\cdot \pi_d^{SSE}(1|21)+1\cdot \pi_d^{SSE}(2|11)}{\pi_d^{SSE}(2|11)+\pi_d^{SSE}(1|21)}=1$. 
Then $D(\pi_d^{ZD}, \pi_a^{BR}(\pi_d^{ZD}),\textbf{1})D(\pi_d^{SSE}, \pi_a^{BR}(\pi_d^{SSE}),\textbf{1} )\leqslant D(\textbf{1})$. 
Thus, for $\lambda\in\Gamma_1$, 
$$
\begin{aligned}
& D(\pi_d^{SSE},\lambda\pi_a^{BR}(\pi_d^{SSE})+(1-\lambda)\pi_a^*,S^d)D(\pi_d^{ZD},\lambda\pi_a^{BR}(\pi_d^{ZD})+(1-\lambda)\pi_a^*,\textbf{1})\\
&- D(\pi_d^{ZD},\lambda\pi_a^{BR}(\pi_d^{ZD})+(1-\lambda)\pi_a^*,S^d)D(\pi_d^{SSE},\lambda\pi_a^{BR}(\pi_d^{SSE})+(1-\lambda)\pi_a^*,\textbf{1})\\
\geqslant & \lambda^4 \left(U_d^{SSE}-U_{12}^d\right)D(\textbf{1})  - A (1-\lambda)^4-B\lambda(1-\lambda). \\
\end{aligned}
$$
For any $\lambda\in\Gamma_1=\{\lambda\in[0,1]|(U_{d}^{SSE}-U_{12}^d)D(\textbf{1})\lambda^4- A(1-\lambda)^4-B\lambda(1-\lambda)\geqslant 0\}$, we have $U_d\left(\pi_d^{SSE}\!,\!\pi_d^{\lambda}(\pi_d^{SSE}\!,\!\pi_d^*)\!\right)\geqslant U_d\left(\pi_d^{ZD}\!,\!\pi_d^{\lambda}(\pi_d^{ZD},\!\pi_d^*)\!\right).$
Moreover, 
$$
\begin{aligned}
&U_d(\pi_d^{SSE},\lambda\pi_a^{BR}(\pi_d^{SSE})+(1-\lambda)\pi_a^*)-U_d(\pi_d^{ZD},\lambda\pi_a^{BR}(\pi_d^{ZD})+(1-\lambda)\pi_a^*)\\
=& \frac{D(\pi_d^{SSE},\lambda\pi_a^{BR}(\pi_d^{SSE})+(1-\lambda)\pi_a^*,S^d)}{D(\pi_d^{SSE},\lambda\pi_a^{BR}(\pi_d^{SSE})+(1-\lambda)\pi_a^*,\textbf{1})}-\frac{D(\pi_d^{ZD},\lambda\pi_a^{BR}(\pi_d^{ZD})+(1-\lambda)\pi_a^*,S^d)}{D(\pi_d^{ZD},\lambda\pi_a^{BR}(\pi_d^{ZD})+(1-\lambda)\pi_a^*,\textbf{1})}\\
=&\frac{1}{C(\pi_d^{ZD},\pi_d^{SSE},\pi_a^*,\lambda)} \left( D(\pi_d^{SSE},\lambda\pi_a^{BR}(\pi_d^{SSE})+(1-\lambda)\pi_a^*,S^d)D(\pi_d^{ZD},\lambda\pi_a^{BR}(\pi_d^{ZD})+(1-\lambda)\pi_a^*,\textbf{1})\right.\\
&\left.-D(\pi_d^{ZD},\lambda\pi_a^{BR}(\pi_d^{ZD})+(1-\lambda)\pi_a^*,S^d)D(\pi_d^{SSE},\lambda\pi_a^{BR}(\pi_d^{SSE})+(1-\lambda)\pi_a^*,\textbf{1})\right)\\
\leqslant & \frac{\left(U_d^{SSE}-U_{12}^d\right)C(\pi_d^{ZD},\pi_d^{SSE},\pi_a^*,1)\lambda^4 - A(1-\lambda)^4+B\lambda(1-\lambda)}{C(\pi_d^{ZD},\pi_d^{SSE},\pi_a^*,\lambda)}\\
=& H(\pi_d^{ZD},\pi_d^{SSE},\pi_a^*,\lambda),
\end{aligned}
$$
where $H(\pi_d^{ZD},\pi_d^{SSE},\pi_a^*,\lambda)$ was defined in (\ref{eq::H}). Therefore, 
$$\begin{aligned}&\min\limits_{\pi_d^{ZD}\in \Xi}\! U_d\left(\pi_d^{SSE}\!,\!\pi_a^{\lambda}(\pi_d^{SSE}\!,\!\pi_a^*)\!\right)\!- \!U_d\left(\pi_d^{ZD}\!,\!\pi_a^{\lambda}(\pi_d^{ZD},\!\pi_a^*)\!\right)\leqslant\!\!\left\{
\begin{array}{ll}
\!\!\!0, & \text{if} \ U_{11}^d\!\geqslant\! U_{22}^d, \ U_{11}^a\!\geqslant \!U_{21}^a,  \\
\!\!\!\!\!H(\pi_d^{ZD},\pi_d^{SSE},\pi_a^*,\lambda), & \text{otherwise}.
\end{array}
\right.
\end{aligned}$$

\section{\label{app::T5}Proof of Theorem 5}
The ZD strategy $\pi_d^{ZD}=\pi^{ZD}(-k, 1, k U_{21}^d -U_{21}^a)$ is feasible for the defender according to Lemma 1. Similar to the analysis in the proof of Theorem 4, we also obtain (\ref{eq::SSEZD-1}) and (\ref{eq::SSEZD-2}). Then
$$
\begin{aligned}
& \left(D(\pi_d^{ZD},\lambda\pi_a^{BR}(\pi_d^{ZD})+(1-\lambda)\pi_a^*,S^d)D(\pi_d^{SSE},\lambda\pi_a^{BR}(\pi_d^{SSE})+(1-\lambda)\pi_a^*,\textbf{1})\right.\\
&\left.- D(\pi_d^{SSE},\lambda\pi_a^{BR}(\pi_d^{SSE})+(1-\lambda)\pi_a^*,S^d)D(\pi_d^{ZD},\lambda\pi_a^{BR}(\pi_d^{ZD})+(1-\lambda)\pi_a^*,\textbf{1})\right)\\
=& \lambda^4\left(\pi_d^{ZD},\pi_a^{BR}(\pi_d^{ZD}),S^d)D(\pi_d^{SSE},\pi_a^{BR}(\pi_d^{SSE}),\textbf{1}) - D(\pi_d^{SSE},\pi_a^{BR}(\pi_d^{SSE}),S^d)D(\pi_d^{ZD},\pi_a^{BR}(\pi_d^{ZD}),\textbf{1}) \right)\\
& + (1-\lambda)^4 \left( D(\pi_d^{ZD},\pi_a^*,S^d)D(\pi_d^{SSE},\pi_a^*,\textbf{1})-D(\pi_d^{SSE},\pi_a^*,S^d)D(\pi_d^{ZD},\pi_a^*,\textbf{1}) \right) +g,
\end{aligned}
$$
where  $g$ is shown in (\ref{eq::gdef}). Since $|g|\leqslant B\lambda(1-\lambda)$ according to (\ref{eq::g}), we have
$$
\begin{aligned}
& D(\pi_d^{ZD},\lambda\pi_a^{BR}(\pi_d^{ZD})+(1-\lambda)\pi_a^*,S^d)D(\pi_d^{SSE},\lambda\pi_a^{BR}(\pi_d^{SSE})+(1-\lambda)\pi_a^*,\textbf{1})
\\
&- 
D(\pi_d^{SSE},\lambda\pi_a^{BR}(\pi_d^{SSE})+(1-\lambda)\pi_a^*,S^d)D(\pi_d^{ZD},\lambda\pi_a^{BR}(\pi_d^{ZD})+(1-\lambda)\pi_a^*,\textbf{1})\\
\geqslant & \lambda^4 \left(U_{12}^d-U_d^{SSE}\right)D(\pi_d^{ZD}, \pi_a^{BR}(\pi_d^{ZD}),\textbf{1})D(\pi_d^{SSE}, \pi_a^{BR}(\pi_d^{SSE}),\textbf{1} )\\
& +(1-\lambda)^4 \left(U_{11}^d-\frac{U_{11}^d \pi_d^{SSE}(1|21)+U_{21}^d\pi_d^{SSE}(2|11)}{\pi_d^{SSE}(2|11)+\pi_d^{SSE}(1|21)} \right)D(\pi_d^{ZD}, \pi_a^*,\textbf{1})D(\pi_d^{SSE}, \pi_a^*,\textbf{1}) \\
&\left.-B\lambda(1-\lambda)\right)\\
\geqslant & \lambda^4 \left(U_{12}^d-U_d^{SSE}\right)D(\textbf{1})  + A (1-\lambda)^4-B\lambda(1-\lambda).
\end{aligned}
$$
As a result, 
$$
\begin{aligned}
&U_d(\pi_d^{ZD},\lambda\pi_a^{BR}(\pi_d^{ZD})+(1-\lambda)\pi_a^*)-U_d(\pi_d^{SSE},\lambda\pi_a^{BR}(\pi_d^{SSE})+(1-\lambda)\pi_a^*)\\
=& \frac{D(\pi_d^{ZD},\lambda\pi_a^{BR}(\pi_d^{ZD})+(1-\lambda)\pi_a^*,S^d)}{D(\pi_d^{ZD},\lambda\pi_a^{BR}(\pi_d^{ZD})+(1-\lambda)\pi_a^*,\textbf{1})}-\frac{D(\pi_d^{SSE},\lambda\pi_a^{BR}(\pi_d^{SSE})+(1-\lambda)\pi_a^*,S^d)}{D(\pi_d^{SSE},\lambda\pi_a^{BR}(\pi_d^{SSE})+(1-\lambda)\pi_a^*,\textbf{1})}\\
=&\frac{1}{C(\pi_d^{ZD},\pi_d^{SSE},\pi_a^*,\lambda)} \left( D(\pi_d^{ZD},\lambda\pi_a^{BR}(\pi_d^{ZD})+(1-\lambda)\pi_a^*,S^d)D(\pi_d^{SSE},\lambda\pi_a^{BR}(\pi_d^{SSE})+(1-\lambda)\pi_a^*,\textbf{1})
\right.\\
&\left.-
D(\pi_d^{SSE},\lambda\pi_a^{BR}(\pi_d^{SSE})+(1-\lambda)\pi_a^*,S^d)D(\pi_d^{ZD},\lambda\pi_a^{BR}(\pi_d^{ZD})+(1-\lambda)\pi_a^*,\textbf{1})\right).
\end{aligned}
$$
Thus, $$U_d(\pi_d^{ZD},\lambda\pi_a^{BR}(\pi_d^{ZD})+(1-\lambda)\pi_a^*)\geqslant U_d(\pi_d^{SSE},\lambda\pi_a^{BR}(\pi_d^{SSE})+(1-\lambda)\pi_a^*),$$ which implies the conclusion.

\section{\label{app::Al}Algorithms}

We show the details of the mentioned algorithms in Applications. Here, we utilize the fictitious play method based on \cite{qiu2021provably} and the Q-learning method based on \cite{li2019cooperation} for the BR strategy of the attacker according to players' action history.

\noindent\begin{minipage}[t]{0.49\linewidth}
\begin{algorithm}[H]
	\caption{Fictitious Play of the Boundedly Rational Attacker}
	\label{al::q_learning}
	\renewcommand{\algorithmicrequire}{\textbf{Input:}}
	\renewcommand{\algorithmicensure}{\textbf{Initialize:}}
	\begin{algorithmic}[1]
		\REQUIRE  Rational factor: $\lambda$, stubborn strategy: $\pi_a^*$.  %%input
		\ENSURE The defender's strategy frequency $\hat{\pi}_d(d|s)=0$ for all $d\in\mathcal{D},s\in\mathcal{S}$, and its average payoff $U_d=0$.
% 		\textcolor{red}{ $\xi_1^{0}=\underset{\omega \in \Theta_1}{\operatorname{argmin}}\left\{ \left\langle \omega, -\sigma_1^{0}
% 		\right\rangle
% 		%\nabla \Psi_1^{*}
% 		+\Psi_{i}(\omega)
% 		\right\} $ }     %%output
		%  \STATE   AAAAA  %\textcolor{blue}{\Comment{initialization}}
		\FOR{$t = 1,2,\cdots$}
		\STATE The defender takes $d_t\sim\pi_d$.
		\STATE The attacker takes $a_t\sim\pi_a$.
		\STATE Reach the state $s(t)$, and players get the payoff $r_d(d_t,a_t)$ and $r_a(d_t,a_t)$.
		\STATE $\hat{\pi}_d(d_t|s_{t-1})=\frac{(t-1)\hat{\pi}_d(d_t|s_{t-1})+1}{t}$.
\STATE$\pi_a=\lambda BR(\hat{\pi}_d)+(1-\lambda)\pi_a^*$.
\STATE$U_d=\sum\limits_{i=1}^t\frac{r_d(d_i,a_i)}{t}$.
		
		\ENDFOR
		
		%\RETURN EEEEE
	\end{algorithmic}
\end{algorithm}
\end{minipage} \
\begin{minipage}[t]{0.49\linewidth}
\begin{algorithm}[H]
	\caption{Q-learning of the Boundedly Rational Attacker}
	\label{al::q_learning}
	\renewcommand{\algorithmicrequire}{\textbf{Input:}}
	\renewcommand{\algorithmicensure}{\textbf{Initialize:}}
	\begin{algorithmic}[1]
		\REQUIRE   Rational factor: $\lambda$, stubborn strategy: $\pi_a^*$,  $\epsilon_1$, and $\epsilon_2$.  %%input
		\ENSURE $Q(s,a)=0$ for $s\in\mathcal{S},a\in\mathcal{A}$, $\bar{r}_a=0$, and  $U_d=0$.
% 		\textcolor{red}{ $\xi_1^{0}=\underset{\omega \in \Theta_1}{\operatorname{argmin}}\left\{ \left\langle \omega, -\sigma_1^{0}
% 		\right\rangle
% 		%\nabla \Psi_1^{*}
% 		+\Psi_{i}(\omega)
% 		\right\} $ }     %%output
		%  \STATE   AAAAA  %\textcolor{blue}{\Comment{initialization}}
		\FOR{$t = 1,2,\cdots$}
		\STATE The defender takes $d_t\sim\pi_d$.
		\STATE The attacker takes $a_t$ with $(1\!-\!\lambda)$-greedy strategy based on  $Q(s(t-1),b)$.
		\STATE Reach the state $s(t)$. Get $r_d(d_t,a_t)$ and $r_a(d_t,a_t)$.
		\STATE $\delta=r_a(d_t,a_t)-\bar{r}_a+\max\limits_{a'}Q(s(t),a')-Q(s(t-1),a_t)$.
		\STATE $Q(s(t-1),a_t)=Q(s(t-1),a_t)+\epsilon_1\delta$.
		 \IF {$Q(s(t-1),a_t)=\max\limits_{b'}Q(s(t-1),a)$}
		     \STATE $\bar{r}_a=(1-\epsilon_2) \bar{r}_a+\epsilon_2\frac{(t-1)\bar{r}_a+r_a(d_t,a_t)}{t}$.
		 \ENDIF
		\STATE$U_d=\sum\limits_{i=1}^t\frac{r_d(d_i,a_i)}{t}$.
		\ENDFOR
		
		%\RETURN EEEEE
	\end{algorithmic}
\end{algorithm}
\end{minipage}

\end{widetext}

\bibliography{apssamp}% Produces the bibliography via BibTeX.

\end{document}